%% file: main.tex
\documentclass[8pt,twocolumn,twoside,usenames,dvipsnames]{article}

\input{preamble.tex}
\input{macros.tex}
\input{todonotes_config.tex}

\usepackage[utf8]{inputenc}
\usepackage[T1]{fontenc}
\usepackage[scaled]{helvet}
\usepackage{lmodern}
\usepackage{microtype}
\usepackage[twoside,%
letterpaper,includeheadfoot,%
layoutsize={8.125in,10.875in},%
layouthoffset=0.1875in,%
layoutvoffset=0.0625in,%
left=38.5pt,%
right=43pt,%
top=43pt,
bottom=32pt,%
headheight=0pt,
headsep=10pt,%
footskip=25pt]{geometry}
\usepackage[plain]{fancyref}
\usepackage[numbers,sort&compress,merge,round]{natbib}
\usepackage[explicit]{titlesec}
\usepackage[labelfont={bf,sf},labelsep=period,figurename=Fig.,font=small]{caption}
\usepackage[english]{babel}
\usepackage[colorlinks=true,allcolors=blue]{hyperref}
\usepackage{authblk}
\usepackage{algorithm}
\usepackage[noend]{algpseudocode}
\usepackage{colortbl}
\usepackage{booktabs}
\usepackage{setspace}

\titleformat{\subsection}[runin]
  {\normalfont\bfseries}
  {\thesubsection.}
  {0.5em}
  {#1. }
  []

  \newcommand{\significancestatement}[1]{\noindent \setlength{\fboxsep}{5pt} \fcolorbox{white}{lightgray}{\parbox{.93\linewidth}{{\bf \small Significance statement.} \small #1}}}
  \newcommand{\abstractwrapper}[1]{#1} 

\newcommand{\bibwrapper}[1]{{\footnotesize #1}}

\title{\vspace{-4.5em}Deterministic networks for probabilistic computing}

\date{}

\author[a]{Jakob Jordan\thanks{To whom correspondence should be addressed. E-mail: j.jordan@fz-juelich.de}}
\author[b,c]{Mihai A. Petrovici}
\author[b]{Oliver Breitwieser}
\author[b]{Johannes Schemmel}
\author[b]{Karlheinz Meier}
\author[a,d,e]{Markus Diesmann}
\author[a]{Tom Tetzlaff}

\affil[a]{Institute of Neuroscience and Medicine (INM-6) and Institute for Advanced Simulation (IAS-6) and JARA-Institute Brain Structure Function Relationship (JBI 1 / INM-10), J\"ulich Research Centre, J\"ulich, Germany}
\affil[b]{Kirchhoff Institute for Physics, Ruprecht-Karls-University Heidelberg, Heidelberg, Germany}
\affil[c]{Department of Physiology, University of Bern, Bern, Switzerland}
\affil[d]{Department of Psychiatry, Psychotherapy and Psychosomatics, Medical Faculty, RWTH Aachen University, Aachen, Germany}
\affil[e]{Department of Physics, Faculty 1, RWTH Aachen University, Aachen, Germany\vspace{-1.25em}}

\begin{document}

\maketitle

\input{significance.tex}

\abstractwrapper{
  \input{abstract.tex}
}
\section*{Introduction}

\input{intro.tex}

\section*{Results}

\input{poolsize.tex}

\input{generativetask.tex}

\input{entropy.tex}

\input{bmsize.tex}

\input{lifsampling.tex}

\section*{Discussion}

\input{discussion.tex}

\subsection*{Acknowledgments}

\input{acks.tex}

\bibwrapper{

\input{main.bbl}
}

\clearpage

\section{Supplementary information}

\input{supplement-methods.tex}
\input{supplement-supp.tex}

\end{document}

%% file: preamble.tex
\usepackage{amsmath}
\usepackage{amsfonts}
\usepackage{amssymb}
\usepackage{graphicx}
\usepackage{textcomp}
\usepackage{soul} 
\usepackage{xr} 
\usepackage[separate-uncertainty=true]{siunitx}
\usepackage[plain]{fancyref}
\usepackage{xcolor}
\usepackage{tabularx}       

\usepackage[final]{changes}

\usepackage{algorithm}


%% file: macros.tex
\definecolor{JJ}{RGB}{200,120,79}
\definecolor{TT}{RGB}{0,200,0}
\definecolor{MAP}{RGB}{255,105,180}
\definecolor{MD}{RGB}{93,140,174} 
\definecolor{OJB}{RGB}{13,181,175}

\definechangesauthor[color=JJ]{JJ}
\definechangesauthor[color=TT]{TT}
\definechangesauthor[color=MAP]{MAP}
\definechangesauthor[color=MD]{MD}
\definechangesauthor[color=OJB]{OJB}

\newcommand{\intrinsic}{intrinsic}
\newcommand{\Intrinsic}{Intrinsic}
\newcommand{\private}{private}
\newcommand{\Private}{Private}
\newcommand{\shared}{shared}
\newcommand{\Shared}{Shared}
\newcommand{\network}{network}
\newcommand{\Network}{Network}

\newcommand{\uV}{\,\mu\text{V}}
\newcommand{\mV}{\,\text{mV}}
\newcommand{\ms}{\,\text{ms}}
\newcommand{\uS}{\,\text{\textmu S}}
\newcommand{\nF}{\,\text{nF}}
\newcommand{\nA}{\,\text{nA}}
\newcommand{\Hz}{\,\text{Hz}}
\newcommand{\kHz}{\,\text{kHz}}

\newcommand{\figlabel}[1]{{\bf #1}}

\renewcommand{\vec}[1]{{\bf #1}}

\newcommand{\betaeff}{\beta_\text{eff}}

\newcommand{\DKL}{D_\text{KL}}

\newcommand{\N}{N}
\newcommand{\Nbm}{M}
\newcommand{\Nbmrec}{m}

\newcommand{\erfc}[1]{\text{erfc}\left(#1\right)}

\newcommand{\Fstoch}[1]{\frac{1}{1+e^{-\beta #1}}}
\newcommand{\Fstochinline}[1]{(1+e^{-\beta #1})^{-1}}

\newcommand{\Ferfc}[1]{\frac{1}{2}\erfc{\frac{#1}{\sqrt{2\sigma^2}}}}
\newcommand{\Ferfcinline}[1]{\erfc{#1/\sqrt{2\sigma^2}}/2}
\newcommand{\sgn}[1]{\text{sign}\left(#1\right)}

\newcommand{\EW}[2][]{\left\langle#2\right\rangle_{#1}}
\newcommand{\mydh}{\mathrm{d}h\;}
\newcommand{\dxi}{\mathrm{d}\xi}
\DeclareMathOperator*{\argmin}{arg\,min}

\newcommand{\h}{h}                                                    

\newcommand{\vs}{\vec{s}}                                             
\newcommand{\zbar}{\langle z \rangle}

\newcommand{\w}{w}
\newcommand{\m}{m}
\newcommand{\mubm}{\mu_\text{BM}}

\newcommand{\K}{K}

\newcommand{\valuebeta}{1}
\newcommand{\valueNbm}{100}
\newcommand{\valueNbmrec}{6}
\newcommand{\valueK}{200}
\newcommand{\valueN}{222}
\newcommand{\valuegamma}{0.3}
\newcommand{\valueT}{10^5}
\newcommand{\valuetrials}{5}
\newcommand{\valuetrialsgen}{20}
\newcommand{\valuea}{2}
\newcommand{\valueb}{2}
\newcommand{\valuew}{0.3}
\newcommand{\valueg}{8}
\newcommand{\valuemu}{0}

\newcommand{\weightscaling}{1/\sqrt{\Nbm}}
\newcommand{\entropy}{S}

\newcommand{\tauref}{\tau_\text{ref}}
\newcommand{\tausyn}{\tau_\text{syn}}
\newcommand{\cmem}{C_\text{m}}
\newcommand{\gleak}{g_\text{L}}
\newcommand{\vm}{V}
\newcommand{\vrest}{V_\text{rest}}
\newcommand{\vreset}{V_\text{reset}}
\newcommand{\vthresh}{V_\text{th}}
\newcommand{\Isyn}{I_\text{syn}}
\newcommand{\nuex}{\nu_\text{ex}}
\newcommand{\nuin}{\nu_\text{in}}

\newcommand{\valueKlif}{1000}
\newcommand{\valueNbmreclif}{10}
\newcommand{\valuetrialslif}{10}

\newlength{\columnwidthleft}
\newlength{\columnwidthmiddle}

\newcommand{\modelhdr}[3]{
  \multicolumn{#1}{|l|}{
    \color{white}
    \cellcolor[gray]{0.0}
    \textbf{\makebox[0pt][l]{#2}\hspace{0.5\textwidth}\makebox[0pt][c]{#3}}
  }
}
\newcommand{\parameterhdr}[3]{
  \multicolumn{#1}{|l|}{
    \color{black}\cellcolor[gray]{0.8}
    \textbf{\makebox[0pt][l]{#2}\hspace{0.5\textwidth}\makebox[0pt][c]{#3}}
  }
}

\newcommand{\fancyrefalgprefix}{alg}
\frefformat{plain}{\fancyrefalgprefix}{algorithm\fancyrefdefaultspacing#1 #2}

%% file: todonotes_config.tex
\usepackage{setspace}
\usepackage[disable]{todonotes}


%% file: significance.tex
\significancestatement{
Probabilistic computing is believed to underly our ability to sense and act in the face of uncertainty. However, implementations of neural-network models performing probabilistic computations on traditional computers or neuromorphic hardware require large numbers of independent noise sources, thereby imposing substantial extra computational load as well as energy and material demands. Furthermore, it remains unclear how appropriate noise is generated in the biological substrate. We demonstrate that the performance of neural networks realizing probabilistic computations is seriously impaired if the noise sources are limited, but that this problem is naturally solved by generating noise with deterministic recurrent neural networks. Computation and noise generation are thereby using the same neural substrate, without the need for dedicated noise-generation infrastructure.
}

%% file: abstract.tex
\begin{abstract}
Neural-network models of high-level brain functions such as memory recall and reasoning often rely on the presence of stochasticity. The majority of these models assumes that each neuron in the functional network is equipped with its own private source of randomness, often in the form of uncorrelated external noise. However, both in vivo and in silico, the number of noise sources is limited due to space and bandwidth constraints. Hence, neurons in large networks usually need to share noise sources. Here, we show that the resulting shared-noise correlations can significantly impair the performance of stochastic network models. We demonstrate that this problem can be overcome by using deterministic recurrent neural networks as sources of uncorrelated noise, exploiting the decorrelating effect of inhibitory feedback. Consequently, even a single recurrent network of a few hundred neurons can serve as a natural noise source for large ensembles of functional networks, each comprising thousands of units. We successfully apply the proposed framework to a diverse set of binary-unit networks with different dimensionalities and entropies, as well as to a network reproducing handwritten digits with distinct predefined frequencies.
Finally, we show that the same design transfers to functional networks of spiking neurons.
\end{abstract}

%% file: intro.tex

The high in-vivo response variability of cortical neurons observed in electrophysiological recordings has often been interpreted in the context of ongoing probabilistic computation \citep{hoyer2003interpreting,berkes2011spontaneous,orban2016neural}.
Probabilistic inference as a principle for brain function has attracted increasing attention over the past decades.
Simultaneously, it has been found that intrinsically stochastic neural networks are a suitable substrate for machine learning \citep{hinton2006reducing,salakhutdinov2009deep}.
Regardless of the purported source of stochasticity \citep{faisal2008noise} -- noise in synaptic transmission \citep{branco2009probability}, ion channel noise \citep{white2000channel} or spiking background input \citep{holt1996comparison} -- these findings have led to the incorporation of noise mechanisms into computational neuroscience models \citep{burkitt2006reviewI,burkitt2006reviewII,destexhe2006neuronal}, in particular to give mechanistic accounts for probabilistic inference in neural substrates \citep{buesing2011neural,petrovici2016stochastic,neftci2016stochastic}.

\begin{figure}[!b]
    \centering
    \includegraphics[width=1.\linewidth]{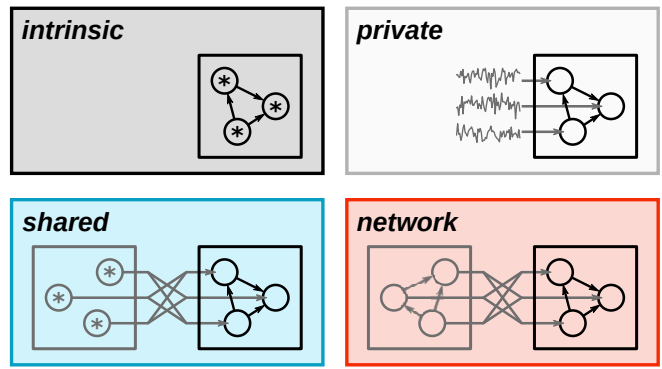}
    \caption{
      Sources of stochasticity (gray) for functional neural networks (black). Stars indicate intrinsically stochastic units. Open circles correspond to deterministic units. \textbf{\Intrinsic}: Intrinsically stochastic units updating their binary states with a probability determined by their total synaptic input. \textbf{\Private}: Deterministic units receiving private additive independent noise. \textbf{\Shared}: Deterministic units receiving noise from a finite population of independent stochastic sources. \textbf{\Network}: Deterministic units receiving quasi-random input generated by a finite recurrent network of deterministic units.
      }
    \label{fig:setups}
\end{figure}

Arguably most widespread is the implementation of noise in neural-network models at the level of individual neurons. 
In this case, neurons are described as intrinsically stochastic units (Fig.~\ref{fig:setups}, \intrinsic) updating their (binary) states in a way that is uniquely determined by their synaptic input \citep{ackley1985learning,buesing2011neural,habenschuss2013stochastic}.
Alternatively, deterministic neurons are equipped with additive private independent noise (Fig.~\ref{fig:setups}, \private), often in the form of Gaussian white noise or random sequences of action potentials (spikes) modeled as Poisson point processes \citep{lundqvist2006attractor,petrovici2016stochastic}.
In either case, each neuron essentially requires its own private pseudorandom number generator.
The implicit assumption of independence of this background noise across units in the network is usually mentioned en passant and goes unchallenged.

\begin{figure}
    \centering
    \includegraphics[width=1.\linewidth]{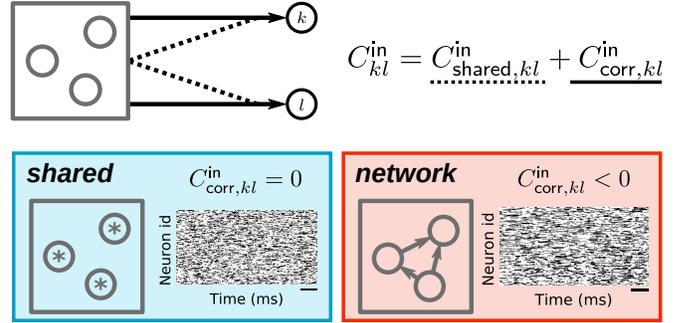}
    \caption{
      Origin of shared-input correlations and their suppression by correlated presynaptic activity. {\bfseries Top:} A pair of neurons $k$ and $l$ (black) receiving input from a finite population of noise sources (gray). The input correlation $C^\text{in}_{kl}$ decomposes into a contribution $C^\text{in}_{\text{shared},kl}$  resulting from shared noise sources (dotted lines) and a contribution $C^\text{in}_{\text{corr},kl}$ (solid lines) due to correlations between sources. If inputs obey Dale's law (they are either excitatory or inhibitory), shared-input correlations are always positive ($C^\text{in}_{\text{shared},kl}>0$). {\bfseries Bottom left:} In the \shared~scenario, sources are by definition uncorrelated ($C^\text{in}_{\text{corr},kl}=0$) and cannot compensate for shared-input correlations. {\bfseries Bottom right:} In inhibition-dominated neural networks (\network~case), correlations between units arrange such that $C^\text{in}_{\text{corr},kl}$ is negative, thereby compensating for shared-input correlations such that $C^\text{in}_{kl} \approx{}0$ \citep{renart2010asynchronous,tetzlaff2012decorrelation}. Raster displays in the bottom row illustrate activity of binary noise sources in the \shared~(left) and the \network~scenario (right). Scale bars correspond to $200\,\ms$.
      }
    \label{fig:intro}
\end{figure}

However, feeding a large amount of uncorrelated noise into a physical realization of a system comes with costs.
In particular, the supply of randomness is physically limited by the input bandwidth of the system, which is necessarily finite for systems that occupy a finite volume, be they biological or synthetic.
This poses a fundamental problem for the biological plausibility of network models of probabilistic inference, as well as for their emulation on physical, neuromorphic devices \citep{indiveri2011neuromorphic,petrovici2014characterization}.
A straightforward solution is to limit the size of the available pool of noise sources (Fig.~\ref{fig:setups}, \shared).
This invariably leads to sharing of noise sources among neurons (Fig.~\ref{fig:intro}), which, in turn, violates the assumption of independence and potentially impairs the performance of the network.

The present work demonstrates that replacing the finite ensemble of independent noise sources by a recurrent neural network (Fig.~\ref{fig:setups}, \network) alleviates the problem of shared noise.
In recurrent neural networks with dominant inhibition, shared-input correlations are dynamically suppressed by a specific correlation structure of network activity \citep{renart2010asynchronous,tetzlaff2012decorrelation}.
We propose to exploit the effect to suppress shared-noise correlations in functional networks resulting from a finite number of noise sources (see Fig.~\ref{fig:intro}).
The mechanism is consistent with biology and simultaneously useful for the implementation of probabilistic computing paradigms on large-scale neuromorphic platforms \citep{schemmel2010wafer,furber2013overview}.

Neural network models derived from Boltzmann machines \cite{ackley1985learning} are representative examples of stochastic models. Such networks are widely used in machine learning \citep{hinton2006reducing,salakhutdinov2009deep}, but also in theoretical neuroscience as models of brain dynamics and function \citep{ginzburg1994theory,buesing2011neural,petrovici2016stochastic}. For the purpose of the present study, the advantage of models of this class lies in our ability to quantify their functional performance when subject to limitations in the quality of the noise sources.

%% file: poolsize.tex

\subsection*{Networks with additive private Gaussian noise approximate Boltzmann machines}

Boltzmann machines \citep[BMs, see][]{ackley1985learning} are symmetrically connected networks of intrinsically stochastic binary units. With an appropriate update schedule and parametrization, the network dynamics effectively implement Gibbs sampling from arbitrary Boltzmann distributions \cite{geman1984gibbs}.
A given network realization leads to a particular frequency distribution of network states. Efficient training methods \citep{hinton2002training,hinton2006reducing} can fit this distribution to a given data distribution by modifying network parameters. In the following we investigate to what extent the functional performance of BM-like stochastic networks is altered if the \intrinsic{} stochasticity assumed in BMs is replaced by \private{}, \shared{} or \network{}-generated additive noise (Fig.~\ref{fig:setups}). If not otherwise indicated, we consider BMs with random connectivity not trained for a specific task. Due to the specific noise-generation processes, the neural network implementations deviate from the mathematical definition of a BM. We therefore refer to these implementations as ``sampling networks''.

In BMs, the intrinsically stochastic units $i \in \{1,\dots,\Nbm\}$ are activated according to a logistic function $F_i(h_i ) = \Fstochinline{h_i}$ of their input field $\h_i = \sum_{j=1}^\Nbm \w_{ij} s_j + b_i$ with inverse temperature $\beta$, synaptic weight $w_{ij}$ between unit $j$ and unit $i$, presynaptic activity $s_j\in\{0,1\}$, and bias $b_i$ (details see SI). In contrast to this intrinsic stochasticity, a more natural model \citep{bryant1976spike,hinton1984boltzmann,mainen1995reliability} considers additive Gaussian noise $\xi_i \sim \mathcal{N}(\mu, \sigma^2)$ on the input to deterministic neurons with Heaviside activation function $F_i(h_i) = \Theta(h_i + \xi_i)$. Essentially normally distributed input naturally emerges in units receiving a large number of inputs from uncorrelated sources. Deterministic units receiving private Gaussian noise resemble units with a probabilistic update rule. In contrast to units in BMs, their effective activation function is a shifted error function $F_i(h_i) = \Ferfcinline{h_i + \mu_i}$.
We minimize the mismatch between the two activation functions by relating the standard deviation $\sigma$ of the Gaussian noise to the inverse temperature $\beta$ (see SI). For a given noise strength, this defines an effective inverse temperature $\betaeff$. To emulate a BM at inverse temperature $\beta$, we rescale all weights and biases: $b_i \rightarrow \beta / \betaeff \, b_i - \mu_i, w_{ij} \rightarrow \beta / \betaeff \, w_{ij}$. The Kullback-Leibler divergence $\DKL(p, p^*)$ between the empirical state distribution $p$ of the sampling network and the state distribution $p^*$ generated by a BM over a subset of $\Nbmrec$ units quantifies the sampling error.

\begin{figure}
  \centering
  \includegraphics[width=.96\linewidth]{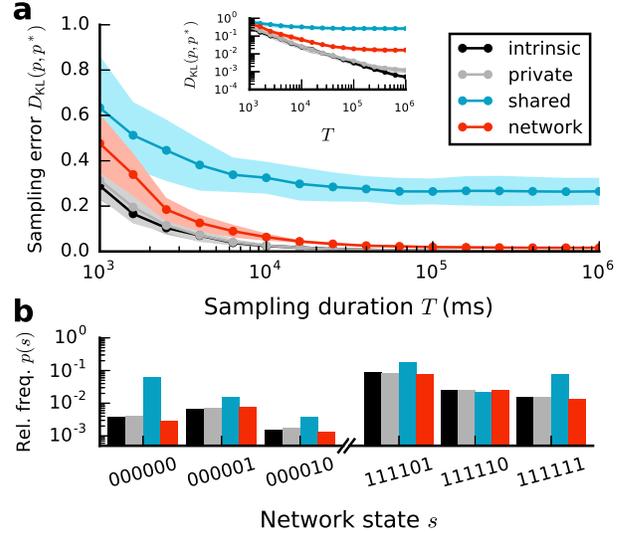}
  \caption{
    (\figlabel{a}) Sampling error as measured by Kullback-Leibler divergence $\DKL(p,p^*)$ between the empirical state distribution $p$ of a functional (sampling) network and the state distribution $p^*$ generated by the corresponding Boltzmann machine as a function of sampling duration $T$ for different sources of stochasticity (legend, cf.~Fig.~\ref{fig:setups}). Error bands indicate mean $\pm$ SEM over $\valuetrials$ random network realizations. Inset: Same data as main panel in double-logarithmic representation. (\figlabel{b}) Relative frequencies (vertical, log scale) of six exemplary states $\vs$ (horizontal) for $T=10^6\ms$. Parameters: $\beta=\valuebeta$, $\Nbm=\valueNbm$, $\K=\valueK$, $N=\valueN$ (see SI).
  }
  \label{fig:results_convergence}
\end{figure}

For matched temperature, networks of deterministic units with additive Gaussian noise closely approximate BMs (Fig.~\ref{fig:results_convergence}, gray vs.~black). The sampling error decreases as a function of the sampling duration $T$, and saturates at a small but finite value (Fig.~\ref{fig:results_convergence}\figlabel{a}, gray) due to remaining differences in the activation functions.
The residual differences between the stationary distributions (Fig.~\ref{fig:results_convergence}\figlabel{b}, black vs.~gray bars) are significantly smaller than the differences in relative probabilities of different network states.

The assumption of private Gaussian noise generated by pseudorandom number generators is hard to reconcile with biology and difficult to achieve in existing neuromorphic devices. In the following, the Gaussian noise is replaced by input from binary and, subsequently, spiking units. As a consequence, the noise of the sampling units exhibits jumps with finite amplitudes determined by the weights of the incoming connections. Only if the number of input events per sampling unit is large and the weights are small, the collective signal resembles Gaussian noise. The sampling error resulting from \private{} Gaussian noise therefore constitutes a lower bound on the error achievable by sampling networks supplied with noise from a finite pool of binary or spiking sources.

\subsection*{Shared-noise correlations impair sampling performance}

Given the space and bandwidth constraints, both in vivo and in silico, neurons in functional networks have to share noise sources to gather random input at a sufficiently high frequency. By replacing private noise with a large number of inputs from a finite pool of independent noise sources, we investigate how strongly the resulting shared-noise correlations distort the sampled distribution of network states. The noise sources are stochastic binary units with an adjustable average activity $\EW{z}$. To achieve a high input event count, each sampling unit is randomly assigned $\K$ inputs. For each unit, these are randomly chosen from a common pool of $\N$ sources. On average, a pair of neurons in the sampling network hence shares $\K^2/\N$ noise sources. The ensemble of noise sources is comprised of $\gamma \N$ excitatory and $(1-\gamma)\N$ inhibitory units, projecting to their targets with weights $w$ and $-gw$, respectively. The input field for a single unit in the sampling network is then given by $h_i' = \sum_{j=1}^M w_{ij} s_j + b_i + \sum_{k=1}^N m_{ik} z_k$, where $m_{ik}$ represents the strength of the connection from the $k$th noise source to the $i$th sampling unit. For homogeneous connectivity, the second term in $h_i'$ can be approximated by a normal distribution with mean $\mu = K w (\gamma - (1 - \gamma)g) \EW{z}$ and variance $\sigma^2 = K w^2 (\gamma + (1 - \gamma) g^2) \EW{z} \rangle (1-\EW{z})$ (details see SI).

\begin{figure}[t]
    \centering
    \includegraphics[width=1.\linewidth]{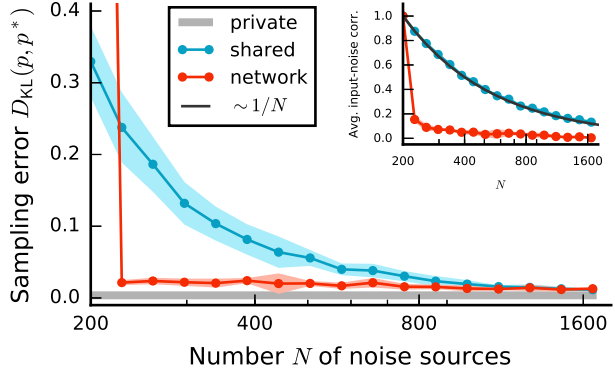}\\
    \caption{
        Sampling error $\DKL(p,p^*)$ as a function of the number $\N$ of noise sources. Error bands indicate mean $\pm$ SEM over $\valuetrials$ random-network realizations. Inset: Dependence of average input correlation coefficient $\rho$ of mutually unconnected sampling units on $\N$. Black curve represents $\sim 1/N$ fit. Sampling duration $T=\valueT\ms$. Remaining parameters as in Fig.~\ref{fig:results_convergence}.
    }
    \label{fig:results_dkl_over_N}
\end{figure}

If $\K \approx \N$, shared-input correlations are large and the sampling error is substantial, even for long sampling duration (Fig.~\ref{fig:results_convergence}, blue curve and bars).
Increasing $\N$ while keeping $\K$ fixed leads to a gradual decrease of shared-input correlations ($\sim 1/\N$) and therefore to a reduction of the sampling error (Fig.~\ref{fig:results_dkl_over_N}, blue curves).
For large $\N \gg \K$, the sampling error approaches values comparable to those obtained with private Gaussian noise (Fig.~\ref{fig:results_dkl_over_N}, blue vs.~gray curves).
For a broad range of $\N$, the sampling error and the average shared-input correlation exhibit a similar trend ($\sim 1/\N$).

\subsection*{Network-generated noise recovers sampling performance}

In recurrent neural networks, inhibitory feedback naturally suppresses shared-input correlations through the emerging activity patterns \citep[Fig.~\ref{fig:intro}; ][]{tetzlaff2012decorrelation}. Here we exploit this effect to minimize the detrimental influence of shared-input correlations that arise due to a limited number of noise sources. To this end, we replace the finite ensemble of independent stochastic sources with a recurrent network of deterministic units (Fig.~\ref{fig:setups}, red). This noise network comprises excitatory and inhibitory binary units with a Heaviside activation function (see SI). Connections from the noise sources to the sampling network follow the same statistics as in the previous section. The connectivity within the noise network is random, sparse and homogeneous \citep{vreeswijk1996chaos,renart2010asynchronous,helias2014intricorr,dahmen16corrfluc}. To achieve optimal suppression of shared-input correlations, connectivity within the noise network needs to be statistically identical to the connectivity between the noise network and the sampling network.
As discussed above, the additional contribution to the input fields $h_i'$ of neurons in the sampling network can be approximated by a normal distribution $\mathcal{N}(\mu,\sigma^2)$. However, an additional term in the input variances arises from correlations between units in the noise network (for details see SI).

Using a recurrent network for noise generation considerably decreases the sampling error compared to the error obtained with a finite number of independent sources, even if the shared-input correlations are substantial (Fig.~\ref{fig:results_convergence}, red vs.~blue curve). Over a large range of noise-network sizes $\N$, the input correlation and, hence the sampling error, are significantly reduced (Fig.~\ref{fig:results_dkl_over_N}, red vs.~blue). In this range, the sampling error is comparable to the error obtained with private Gaussian noise and almost independent of $\N$ (Fig.~\ref{fig:results_dkl_over_N}, red vs.~gray). Only if the noise network becomes too dense ($\K \approx \N$), its dynamics lock into a fixed point and the sampling performance breaks down.

%% file: generativetask.tex
\subsection*{%
  Deterministic neural networks serve as a suitable noise source for a model of handwritten-digit generation%
}
\begin{figure}[t]
    \centering
    \includegraphics[width=1.\linewidth]{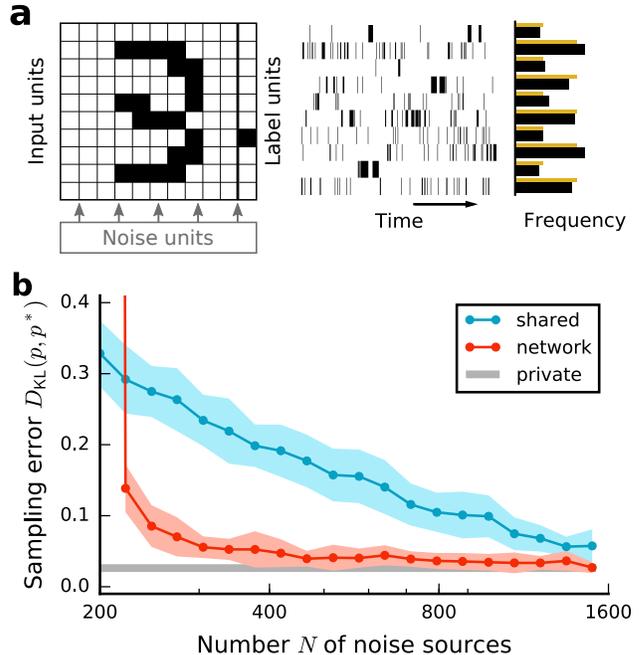}
    \caption{
        Performance of a generative network trained on an imbalanced subset of the MNIST dataset for different noise sources. (\figlabel{a}) Left: Sketch of the network consisting of external noise inputs, input units, trained to represent patterns corresponding to handwritten digits, and label units trained to indicate the currently active pattern. Right: Network activity and trial-averaged relative activity of label units for \intrinsic{} noise (black) and target distribution (yellow), with even digits occurring twice as often as odd digits. (\figlabel{b}) Sampling error $\DKL(p, p^*)$ between the empirical state distribution $p$ of label units and the state distribution $p^*$ of label units generated by the corresponding Boltzmann machine as a function of the number $\N$ of noise sources for \shared{} and \network{} case. Error bands indicate mean $\pm$ SEM over $\valuetrialsgen$ trials with different initial conditions and noise realizations.
    }
    \label{fig:results_RBM}
\end{figure}

All realizations within the ensemble of unspecific, randomly generated sampling networks considered so far exhibit consistent performance characteristics (cf.~narrow error bands in Fig.~\ref{fig:results_convergence} and Fig.~\ref{fig:results_dkl_over_N}). We obtain similar behavior for a sampling network where the weights and biases are not chosen randomly but trained for a specific task -- the generation of handwritten digits with imbalanced class frequencies (see SI): (i) networks with \private{} external noise perform close to optimal, (ii) \shared{} noise correlations impair network performance, and (iii) the performance is restored by employing a recurrent \network{} for noise generation (Fig.~\ref{fig:results_RBM}). Thus deterministic recurrent neural networks qualify as a suitable noise source for practical applications of neural networks performing probabilistic computations.

%% file: entropy.tex
\subsection*{Shared-input correlations impair network performance for high-entropy tasks}

The dynamics of a BM representing a high-entropy distribution evolve on a flat energy landscape with shallow minima. Here, the sampling process is sensitive to perturbations in statistical dependencies, such as those caused by shared-input correlations. In contrast, the sampling dynamics in BMs representing low-entropy distributions with pronounced peaks are dominated by deep minima in the energy landscape. In this case, noise correlations have little effect.

\begin{figure}[t]
  \centering
  \includegraphics[width=1.\linewidth]{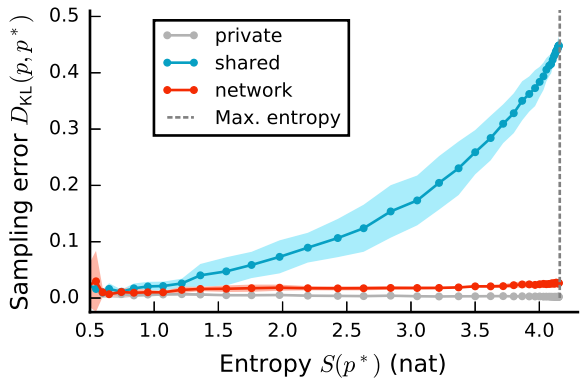}
  \caption{
    Sampling error $\DKL(p,p^*)$ as a function of the entropy $\entropy$ of the target distribution. Error bands indicate mean $\pm$ SEM over $\valuetrials$ random network realizations. Vertical dashed gray line indicates maximal entropy, corresponding to a uniform target distribution. Sampling duration $T=\valueT\ms$. Remaining parameters as in Fig.~\ref{fig:results_convergence}.
  }
  \label{fig:results_dkl_over_beta}
\end{figure}

Fig.~\ref{fig:results_dkl_over_beta} systematically varies the entropy of the target distribution by changing the inverse temperature $\beta$ in a BM and adjusting the relative noise strength in the other cases accordingly. Since $\beta$ always appears as a multiplicative factor in front of weights and biases, this is equivalent to scaling weights and biases globally. For small entropies, the sampling error for \shared{} and \network{} noise is comparable to the error obtained with \private{} noise, despite substantial shared-input correlations. Consistent with the intuition provided above, the sampling error for \shared{} noise increases significantly with increasing entropy, whereas in the other cases it remains low.

We conclude that generally the effect of shared-noise correlations on the functional performance of sampling networks depends on the entropy of the target distribution. For high-entropy tasks, such as pattern generation, shared-input correlations can be highly detrimental. For low-entropy tasks, such as pattern classification, they play a less significant role. Nevertheless, independent of the entropy of the task, functional performance for network-generated noise is close to optimal.

%% file: bmsize.tex
\subsection*{Small recurrent networks provide large sampling networks with noise}

To achieve a good sampling performance, both the number $\N$ of noise sources as well as the number $\K$ of noise inputs per functional unit need to be sufficiently large (Fig.~\ref{fig:results_dkl_over_N}). Therefore, a certain minimal amount of resources have to be reserved for noise generation. Once these resources are allocated, small recurrent networks can provide noise for large sampling networks without sacrificing computational performance. We note in passing that a single noise network can supply an arbitrary number of independent functional networks with noise.

\begin{figure}[t]
  \centering
  \includegraphics[width=1.\linewidth]{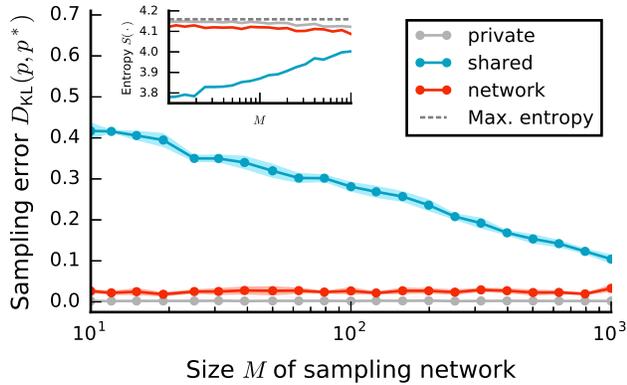} 
  \caption{
    Sampling error $\DKL(p,p^*)$ as a function of the sampling-network size $\Nbm$. Error bands indicate mean $\pm$ SEM over $\valuetrials$ random network realizations. Inset: Entropy of the sampled state distribution $p$ as a function of $\Nbm$. Horizontal dashed dark gray line indicates entropy of uniform distribution, i.e., maximal entropy. Average weight in sampling networks: $\mu_\text{BM}=-0.15/\sqrt{\Nbm}$. Sampling duration $T=\valueT\ms$. Remaining parameters as in Fig.~\ref{fig:results_convergence}.
  }
  \label{fig:results_dkl_over_M}
\end{figure}

Fig.~\ref{fig:results_dkl_over_M} varies the size of the sampling network $\Nbm$, while keeping $\N$ and the number $\Nbmrec$ of observed neurons fixed. By increasing $\Nbm$, the entropy of the marginal distribution over the subset of observed neurons changes (see SI), thereby influencing the sampling performance in the presence of shared-noise correlations. Scaling the weights in the sampling network with $\weightscaling$ \citep{van1998chaotic,renart2010asynchronous,helias2014intricorr} keeps the entropy of the marginal target distribution approximately constant (Fig.~\ref{fig:results_dkl_over_M} inset, gray curve).
In the presence of \private{} noise, the sampling error is small and independent of $\Nbm$ (Fig.~\ref{fig:results_dkl_over_M}). As before, the performance is considerably impaired for \shared{} noise.
The decrease in the error for larger sampling networks cannot be traced back to a change in entropy, by virtue of the weight scaling. Instead, the decrease results from a more efficient suppression of external correlations within the sampling network arising from the growing negative feedback for increasing $\Nbm$ in sampling networks with net recurrent inhibition \citep{helias2014intricorr}. Still, even for large $\Nbm$, the error remains significantly larger than the one obtained with \private{} noise. For \network{} noise, in contrast, the error is almost as small as for \private{} noise, and independent of $\Nbm$. Qualitatively similar findings are also obtained without scaling synaptic weights unless the entropy of the target distribution is too small (see SI).

%% file: lifsampling.tex
\subsection*{Networks of spiking neurons implement neural sampling without noise}

\begin{figure}[t]
  \centering
  \includegraphics[width=1.\linewidth]{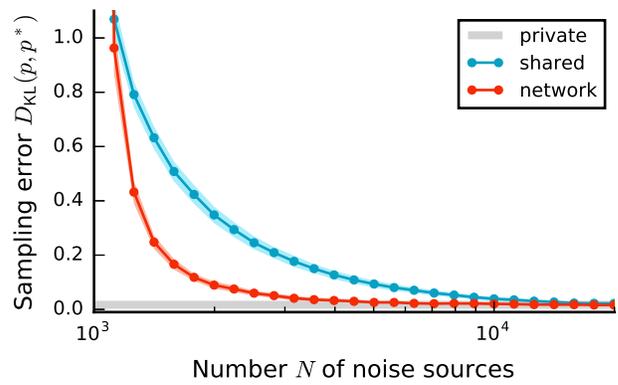}
  \caption{
    Sampling in spiking networks with biologically plausible noise networks. Kullback-Leibler divergence $\DKL(p,p^*)$ between the empirical state distribution $p$ of a sampling network of spiking neurons and the state distribution $p^*$ generated by the corresponding Boltzmann machine as a function of the number $\N$ of noise sources. Error bands indicate mean $\pm$ SEM over $\valuetrialslif$ random network realizations. See SI for model parameters.
  }
  \label{fig:results_lif}
\end{figure}

The results so far rest on networks of binary model neurons. Their dynamics are well understood \citep{ginzburg1994theory,van1998chaotic,renart2010asynchronous,helias2014intricorr,dahmen16corrfluc}, and their mathematical tractability simplifies the calibration of sampling-network parameters for network-generated noise. Neurons in mammalian brains and in neuromorphic hardware communicate, however, predominantly via short electrical pulses (spikes) \citep{nawrocki2016mini,furber2016large}. Indeed, networks of spiking neurons with private external noise can approximately represent arbitrary Boltzmann distributions, if binary-unit parameters are properly translated to spiking-neuron parameters \citep[][ see also gray curve in Fig.~\ref{fig:results_lif}; details see SI]{petrovici2016stochastic,probst2015probabilistic}. Consistent with our results on binary networks, the sampling performance decreases in the presence of shared-noise correlations, but recovers for noise provided by a recurrent network of spiking neurons resembling a local cortical circuit with natural connection density and activity statistics (Fig.~\ref{fig:results_lif}).
Similar to binary networks, a minimal noise-network size $\N$ ensures an asynchronous activity regime, a prerequisite for good sampling performance.
Spiking noise networks that are too densely connected ($\K/\N \rightarrow 1$) tend to synchronize, causing large sampling errors (see red curve in Fig.~\ref{fig:results_lif}).

%% file: discussion.tex


Consistent with the high variability in the activity of biological neural networks \citep{faisal2008noise}, many models of high-level brain function rely on the presence of some form of stochasticity. We propose that additive input from deterministic recurrent neural networks serves as a well controllable source of noise for functional network models. The article demonstrates that networks of deterministic units with input from such noise-generating networks can approximate a large variety of target distributions and perform well in generative tasks. This scheme covers both networks of binary and networks of spiking neurons, and leads to an economic usage of resources in biological and artificial neuromorphic systems.

For conceptual simplicity, the study segregates a neuronal network into a functional and a noise-generating module. In biological substrates, these two modules may be intermingled. Alternatively, a brain may use one sampling network as the noise source for another. In this view, one network's function is another network's noise.

We show that shared-noise correlations can be highly detrimental for sampling from given target distributions.
Generating noise with recurrent neural networks overcomes this problem by exploiting active decorrelation in networks with inhibitory feedback \citep{renart2010asynchronous,tetzlaff2012decorrelation}.
As an alternative solution, the effect of shared-input correlations could be mitigated by training functional network models in the presence of these correlations \citep{bytschok2017spike}.
However, this approach is specific to particular network models.
Moreover, it prohibits porting of models between different substrates.
Networks previously trained under specific noise conditions will not perform well in the presence of noise with a different correlation structure. 
Our approach, in contrast, constitutes a general-purpose solution which can also be employed for models that cannot easily be adapted to the noise statistics, such as hard-wired functional network models \citep[e.g.,][]{lundqvist2006attractor,jonke2016solving} or bottom-up biophysical neural-network models \citep[e.g.,][]{potjans2012cell}.

In biological neural networks, the probabilistic gating of ion channels in the cell membrane \citep{white2000channel} and the variability in synaptic transmission \citep{branco2009probability} constitute alternative potential sources of stochasticity.
However, for the majority of stochastic network models, ion-channel noise is too small to be relevant: in the absence of (evoked or spontaneous) synaptic input, fluctuations in membrane potentials recorded in vitro are in the $\uV$ range and hence negligible compared to the $\mV$ fluctuations necessary to support sampling-based approaches \citep{petrovici2016stochastic}.
Synaptic stochasticity comes in two distinct forms: spontaneous release and variability in evoked postsynaptic response amplitudes.
The rate of spontaneous synaptic events measured at the soma of the target neuron is in the range of a few events per second \citep{hardingham1998reliability,locke1999miniature}. The resulting fluctuations in the input are therefore negligible. The variability in postsynaptic response amplitudes, in contrast, is substantial \citep{branco2009probability}, and has often been suggested as a plausible noise resource for computations in neural circuits \citep{levy2002energy,rosenbaum2012short,maass2014noise,kappel2015network,neftci2016stochastic,muller2017neural}. Due to its multiplicative, state-dependent nature, this form of noise is fundamentally different from the noise usually employed in sampling models.
Its role for approximate inference in neural substrates remains unclear. 

Some neuromorphic-hardware systems follow alternative approaches to the generation of uncorrelated noise for stochastic network models, such as exploiting thermal noise and trial-to-trial fluctuations in neuron parameters \citep[see, e.g.,][]{hamid2011probabilistic,binas2016spiking,sengupta2016magnetic}. 
However, hardware systems need to be specifically designed for a particular technique and sacrifice chip area that otherwise could be used to house neurons and synapses. The solution proposed in this article does not require specific hardware components for noise generation. It solely relies on the capability of emulating recurrent neural networks, the functionality most neuromorphic-hardware systems are designed for.
On the neuromorphic system Spikey \citep{pfeil2013six}, for example, it has already been demonstrated that decorrelation by inhibitory feedback is effective and robust, despite large heterogeneity in neuron and synapse parameters and without the need for time-consuming calibrations \citep{pfeil2016effect}. While a full neuromorphic-hardware implementation of the framework proposed here is still pending, the demonstration on Spikey shows that our solution is immediately implementable and feasible.


%% file: acks.tex
This research was supported by the Helmholtz Association portfolio theme SMHB, the J\"ulich Aachen Research Alliance (JARA) and EU Grants 269921 (BrainScaleS), \#604102 and \#720270 (Human Brain Project).
The authors acknowledge support by the state of Baden-W\"urttemberg through bwHPC and the German Research Foundation (DFG) through grant no INST 39/963-1 FUGG.
All spiking network simulations carried out with NEST \citep[][\href{http://www.nest-simulator.org}{http://www.nest-simulator.org}]{gewaltig2007nest}.

%% file: supplement-methods.tex
\subsection{Binary network simulation}
\label{sec:supp-binary-network-sim}

Sampling networks consist of $\Nbm$ binary units that switch from the inactive ($0$) to the active ($1$) state with a probability $F_i(h_i):=p(s_i=1|h_i)$, also referred to as the ``activation function''.
The input field $h_i$ of a unit depends on the state of the presynaptic units and is given by:
\begin{align}
  \h_i(\vs) = \sum_j \w_{ij} s_j + b_i \, .
  \label{eq:supp-input-field}
\end{align}
Here $w_{ij}$ denotes the weight of the connection from unit $j$ to unit $i$ and $b_i$ denotes the bias of unit $i$.
We perform an event-driven update, drawing subsequent inter-update intervals $\tau_i \sim \text{Exp}(\lambda)$ for each unit from an exponential distribution with rate $\lambda := 1/\tau$ with an average update interval $\tau$.
Starting from $t=0$, we update the neuron with the smallest update time $t_i$, choose a new update time for this unit $t_i + \tau_i$ and repeat this procedure until any $t_i$ is larger than the maximal simulation duration $T_\text{max}$.

\subsection{Random sampling networks}
\label{sec:supp-random-sampling-networks}

Weights are randomly drawn from a beta distribution $\text{Beta}(a, b)$ and shifted to have mean $\mubm$.
Weights are symmetric ($w_{ij} = w_{ji}$) and self connections are absent ($w_{ii} = 0$).
To control the average activity in the network, the bias for each unit is chosen such that on average, it cancels the input from the other neurons in the network for a desired average activity $\EW{s}$: $b_i = \Nbm \mu \EW{s}$ \citep{helias2014intricorr}.
Whenever a unit is updated, the state of (a subset) of all units in the sampling network is recorded.
To remove the influence of initial transients, i.e., the burn-in time of the Markov chain, samples during the initial interval of each simulation ($T_\text{warmup}$) are excluded from the analysis.
From the remaining samples we compute the empirical distribution $p$ of network states.
The following sections introduce the activation function for the units for different ways of introducing noise to the system.

\subsection{Intrinsic noise}
\label{sec:supp-intrinsic-noise}

Intrinsically stochastic units switch to the active state with probability
\begin{align}
  F_i(h_i) = \frac{1}{1 + e^{-\beta h_i}} \, ,
  \label{eq:supp-logistic}
\end{align}
where $\beta$ determines the slope of the logistic function and is also referred to as the ``inverse temperature''.
For small $\beta$, changes in the input field have little influence of the update probability, while for large beta a unit is very sensitive to changes in $h_i$ and in the limit $\beta \rightarrow \infty$ the activation function becomes a Heaviside step function.
Symmetric networks with these single-unit dynamics and the update schedule described in Sec.~\ref{sec:supp-binary-network-sim} are identical to Boltzmann machines, leading to a stationary distribution of network states of Boltzmann form:
\begin{align}
  p(\vs)\sim \exp\left(\frac{\beta}{2} \sum_{i,j}\w_{ij}s_is_j + \beta \sum_i b_i s_i\right) \, .
  \label{eq:supp-bm-dist}
\end{align}
Instead of directly prescribing a stochastic update rule like Eq.~\ref{eq:supp-logistic}, we can view these units as deterministic units with a Heaviside activation function and additive noise on the input field:
\begin{align*}
  F_i(h_i) = \Theta(h_i + \xi_i) \; ,
\end{align*}
with $\xi_i \sim 1/2(1-\tanh^2(\xi_i))$ \citep{coolen2001statistical} and $\Theta$ denoting the Heaviside step function
\begin{align}
  \Theta(x) = \begin{cases}
    1 & \text{if } x \ge 0 \\
    0 & \text{else}
  \end{cases}
\end{align}
Averaging over the noise $\xi_i$ yields the probabilistic update rule (Eq.~\ref{eq:supp-logistic}).
However, on biophysical grounds it is difficult to argue for this particular distribution of the noise.

\subsection{Private noise}
\label{sec:supp-private-noise}

We consider a deterministic model in which we assume a more natural distribution for the additive noise, namely Gaussian form ($\xi_i \sim \mathcal{N}(\mu_i, \sigma_i^2)$), for example arising from a large number of independent background inputs \citep{hinton1984boltzmann}.
In this case, the noise averaged activity for fixed $h_i$ is given by:
\begin{align}
  F_i(h_i) =& \int_{-\infty}^\infty \dxi_i \; \Theta(h_i+\xi_i) p(\xi_i) \notag \\
  =& \int_{-h_i}^\infty \dxi_i \; \mathcal{N}(\mu_i, \sigma_i^2) \notag \\
  =& \frac{1}{2}\erfc{-\frac{h_i + \mu_i}{\sqrt{2}\sigma_i}} \;.
     \label{eq:supp-actf-gauss}
\end{align}
Similar to the intrinsically stochastic units (Sec.~\ref{sec:supp-intrinsic-noise}), the update rule for deterministic units with Gaussian noise is effectively probabilistic.
Both functions, share some general properties (bounded, monotonic):
\begin{align*}
  \lim_{h_i \rightarrow -\infty} F_i(h_i) =& 0 \, , \\
  \lim_{h_i \rightarrow \infty} F_i(h_i) =& 1 \, , \\
  \partial_{h_i} F_i(h_i) >& 0\; \forall h_i \, ,
\end{align*}
and one can hence hope to approximate the dynamics in Boltzmann machines with a network of deterministic units with Gaussian noise by a systematic matching of parameters.

\begin{figure}[t]
  \centering
  \includegraphics[width=1.\linewidth]{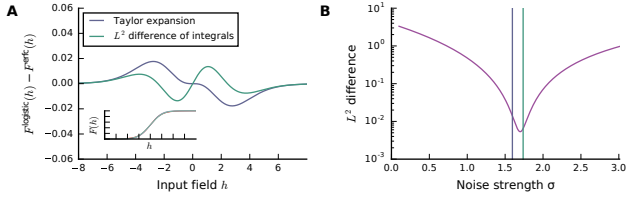}
  \caption{Fit of error function to logistic function via Taylor expansion (purple) and $L^2$ difference of integrals (green). \figlabel{A} Difference of logistic activation function and error function with adjusted $\sigma$ via Eq.~\ref{eq:supp-sigma-beta-taylor} (purple) and via Eq.~\ref{eq:supp-sigma-beta-integral} (green). Inset: activation functions. \figlabel{B} $L^2$ difference of activation functions (Eq.~\ref{eq:supp-l2-norm-actf}) as a function of the strength of the Gaussian noise $\sigma$. Vertical bars indicate $\sigma$ obtained via the respective method.}
  \label{fig:noise-actf-error}
\end{figure}

One approach is to choose parameters for the Gaussian noise such that the difference between the two activation functions is minimized.
To simplify notation we drop the index $i$ in the following calculations.
Since both activation functions are symmetric around zero, we require that their value at $h=0$ is identical, fixing one parameter of the noise distribution ($\mu=0$).
To find an expression for the noise strength $\sigma$, the simplest method equates the coefficients of a Taylor expansion up to linear order of both activation functions around zero.
For the logistic activation function (Eq.~\ref{eq:supp-logistic}) this yields:
\begin{align*}
  F(h) =& 0.5 + 0.25 \beta h + \mathcal{O}(h^2) \, ,
\end{align*}
while for the units with Gaussian noise (Eq.~\ref{eq:supp-actf-gauss}) we obtain
\begin{align*}
  F(h) =& 0.5 + \frac{1}{\sqrt{2\pi}\sigma}h + \mathcal{O}(h^2) \, .
\end{align*}
Equating the coefficients of $h$ gives an expression for the noise strength $\sigma$ as a function of the inverse temperature $\beta$:
\begin{align}
  \sigma(\beta) = \frac{2 \sqrt{2}}{\sqrt{\pi} \beta} \, .
  \label{eq:supp-sigma-beta-taylor}
\end{align}
While this approach is conceptually simple, the Taylor expansion around zero leads to large deviations between the activation functions for input fields different from zero (Fig.~\ref{fig:noise-actf-error}).

Another option taking into account all possible values of $h$ is to minimize the $L^2$ difference of the two activation functions:
\begin{align}
  \sigma =& \argmin_{\sigma'} \int \mydh (l(h) - g(h, \sigma'))^2 \, ,
  \label{eq:supp-l2-norm-actf}
\end{align}
where $l$ denotes the logistic and $g$ the activation function for Gaussian noise.
Since it is not possible to analytically evaluate the resulting integral, we opt for a slightly simpler approach: minimizing the $L^2$ difference of integrals of the activation function from $-\infty$ to $0$:
\begin{align*}
  \sigma =& \argmin_{\sigma'} \left(L(h) \bigg\rvert_{-\infty}^0 - G(h, \sigma') \bigg\rvert_{-\infty}^0\right)^2 \, ,
\end{align*}
with capital letters denoting antiderivatives.
To find the minimal $\sigma$, we take the derivative of the right hand side with respect to $\sigma'$ and equate it with zero:
\begin{align*}
  -2 \left( F(h) \bigg\rvert_{-\infty}^0 - G(h) \bigg\rvert_{-\infty}^0 \right) \partial_\sigma G(h) \bigg\rvert_{-\infty}^0 = 0 \, .
\end{align*}
From this we observe that
\begin{align}
  \left(F(h) \bigg\rvert_{-\infty}^0 - G(h) \bigg\rvert_{-\infty}^0\right) = 0 \, ,
  \label{eq:supp-suff-cond}
\end{align}
is a sufficient condition to satisfy this equation.
We compute the integral of both activation functions.
For the logistic activation function (Eq.~\ref{eq:supp-logistic}) we obtain:
\begin{align*}
  \int \mydh F(h) =& \int \mydh \frac{1}{1 + e^{-\beta h}} \\
  =& h + \frac{\log(1 + e^{-\beta h})}{\beta} \, ,
\end{align*}
with the definite integral
\begin{align*}
  \int_{-\infty}^0 \mydh F(h) =& \frac{\log 2}{\beta} \, ,
\end{align*}
since the two diverging terms for $h \rightarrow -\infty$ cancel.
For the activation function with Gaussian noise (Eq.~\ref{eq:supp-actf-gauss}) we get:
\begin{align*}
  \int \mydh F(h) =& \int \mydh \frac{1}{2}\erfc{\frac{-h}{\sqrt{2}\sigma}} \\
  =& \frac{\sigma}{\sqrt{2 \pi}} e^{- \frac{h^{2}}{2 \sigma^{2}}} + 0.5 h \erfc{\frac{-h}{\sqrt{2} \sigma}} \, ,
\end{align*}
and computing the definite integral leads to:
\begin{align*}
  \int_{-\infty}^0 \mydh F(h) =& \frac{\sigma}{\sqrt{2 \pi}} \, ,
\end{align*}
since the second term vanishes for $h \rightarrow -\infty$ as the complementary error function decreases faster than $|h|^{-1}$.
From Eq.~\ref{eq:supp-suff-cond} we hence find $\sigma$ as a function of $\beta$:
\begin{align}
  \sigma(\beta) = \frac{\log 2 \sqrt{2 \pi}}{\beta} \, .
  \label{eq:supp-sigma-beta-integral}
\end{align}
Even though this value is not minimizing the $L^2$ difference, it provides a better fit than that obtained by simply Taylor expanding around zero, since in this case we are also taking into account the mismatch for larger absolute values of $h$ (Fig.~\ref{fig:noise-actf-error}).
We will hence use Eq.~\ref{eq:supp-sigma-beta-integral} to translate between the inverse temperature $\beta$ of the logistic activation function and the strength $\sigma$ of the Gaussian noise.

\subsection{Shared noise}
\label{sec:supp-shared-noise}

In the previous section we have assumed that each deterministic unit in the sampling network receives private, uncorrelated Gaussian noise.
Now we instead consider a second population $\mathcal{B}$ of $N = |\mathcal{B}|$ mutually unconnected, intrinsically stochastic units with logistic activation functions (cf.~Sec.~\ref{sec:supp-intrinsic-noise}) that provide additional input to units in the sampling network.
In the following we will denote the population/set of units in the sampling network by $\mathcal{S}$ and refer to the second population as the background population or noise population.
The input field for a unit $i$ in the sampling network $\mathcal{S}$ hence contains an additional term arising from projections from the background population (cf.~Eq.~\ref{eq:supp-input-field}):
\begin{align}
  \label{eq:noise-shared-noise}
  h_i' = \underbrace{\sum_{j\in \mathcal{S}} \w_{ij} s_j + b_i}_{h_i} \underbrace{+ \sum_{k\in \mathcal{B}} \m_{ik} z_k}_{\text{background input}} \, .
\end{align}
Here $z_k$ denotes the state of the $k$th unit in the background population $\mathcal{B}$ and $\m_{ij}$ the weight from unit $j$ in the background population to unit $i$ in the sampling network.
Given the total input field $h_i'$, the neurons in the sampling network change their state deterministically, according to
\begin{align}
  F_i(h_i') = \Theta(h_i') \, .
  \label{eq:supp-actf-det}
\end{align}
Since the units in the background population are mutually unconnected, their average activity $\langle z_i \rangle$ can be arbitrarily set by adjusting their bias: $b_k = F^{-1}(\langle z_k \rangle)$, where $F^{-1}$ denotes the inverse of the logistic activation function:
\begin{align*}
  F^{-1}(\zbar) = \frac{1}{\beta}\log \frac{1}{\frac{1}{\zbar} - 1} \, .
\end{align*}
Ignoring the actual state of the background population, we can employ the central limit theorem and approximate the background input in the input field $h_i'$ by a normal distribution with mean and variance given by
\begin{align}
  \mu_i =& \sum_{k\in \mathcal{B}} \m_{ik} \langle z_k \rangle \, ,\\
  \sigma_i^2 =& \sum_{k\in \mathcal{B}} \m_{ik}^2 \langle z_k \rangle (1 - \langle z_k \rangle) \, .
\end{align}
The total input field can then be written as $h_i' = h_i + \xi_i$ with $\xi_i \sim \mathcal{N}(\mu_i, \sigma_i^2)$, as in the case of private uncorrelated Gaussian noise.
However, note that correlations in input fields $h_i'$ and $h_j'$ in the sampling network arise due to units in the background population projecting to multiple units in the sampling network ($\langle (\xi_i -\mu_i) (\xi_j - \mu_j) \rangle$ does not necessarily vanish for all $i, j \in \mathcal{S}$).

For the connections from the background population we use fixed weights and impose Dale's law, i.e., units are either excitatory $\m_{ij} = w > 0 \; \forall i$ or inhibitory $\m_{ij} = -gw < 0 \; \forall i$, with a ratio of excitatory units of $\gamma = |\mathcal{B}_E| / |\mathcal{B}|$.
Here $w \in \mathbb{R}^+$ denotes the excitatory synaptic weight and $g \in \mathbb{R}^+$ a scaling factor for the inhibitory weights.
Each unit $i \in \mathcal{S}$ in the sampling network receives exactly $K = \epsilon N$ inputs from units in the background population.
$\epsilon = \K/\N \in [0, 1]$ is referred to as the connectivity.
We do not allow multiple connections between a unit in the sampling network and unit in the background population.
Assuming all units in the background population have identical average activity $\zbar$, all units in the sampling network receive statistically identical input and, in addition, the equations for the mean and variance simplify to
\begin{align}
  \mu =& K w (\gamma - (1 - \gamma)g)\zbar \, , \\
  \sigma^2 =& K w^2 (\gamma + (1 - \gamma) g^2) \zbar(1-\zbar) \, .
\end{align}
We can hence employ the same procedure as in the previous section to relate the strength of the background input to the inverse temperature of a Boltzmann machine.

\subsection{Network noise}
\label{sec:supp-network-noise}

We now consider a background population of deterministic units projecting to the sampling network.
The background population has sparse, random, recurrent connectivity with a fixed indegree.
Connections in the background population are realized with the same indegrees $K$, weights $w$ and $-gw$ and ratio of excitatory inputs $\gamma$ as the connections to the sampling network (cf.~Sec.~\ref{sec:supp-shared-noise}).
The connection matrix of the background population is hence generally asymmetric.
As before, we can approximate the additional contribution to the input fields of neurons in the sampling network with a normal distribution, with parameters
\begin{align}
  \label{eq:supp-mu-network}
  \mu_i =& \sum_{k\in \mathcal{B}} \m_{ik} \EW{z_k} \, , \\
  \sigma_i^2 =& \sum_{k\in \mathcal{B}} \m_{ik}^2 \EW{z_k} (1 - \EW{z_k}) + \sum_{k\ne l} \m_{ik}\m_{il} c_{kl} \, ,
  \label{eq:supp-sigma-network}
\end{align}
where the additional term in the input variances arises from correlations $c_{kl}:=\EW{(z_k - \EW{z_k})(z_l - \EW{z_l})}$ between units in the background population.
As in the sampling network we choose the bias to cancel the expected average input from other units in the network for a desired mean activity $\EW{z_k}$.
However since the second population exhibits rich dynamics due to its recurrent connectivity the actual average activity will deviate from this value, in particular due to an influence of correlations on the mean activity.
We employ an iterative meanfield-theory approach that allows us to compute average activities and average correlations approximately from the statistics of the connectivity.
We now shortly summarize this approach following \citep{helias2014intricorr}.
Note that in the literature a threshold variable $\theta_i$ is often used instead the bias $b_i$, which differs in the sign: $b_i = -\theta_i$.

For a network of binary units, the joint distribution of network states $p(\vs)$ contains all information necessary to statistically describe the network activity, in particular mean activities and correlations.
It can be obtained by solving the Master equation of the system, which determines how the probability masses of network states evolve over time in terms of transition probabilities between different states \citep{kelly2011reversibility}
\begin{align}
  \partial_t p(\vs_i) = \sum_j p(\vs_i|\vs_j)p(\vs_j) - p(\vs_j|\vs_i)p(\vs_i) \, .
  \label{eq:noise-master-eq}
\end{align}
The first term describes probability mass moving into state $i$ from other states $j$ and the second term probability mass moving from state $i$ to other states $j$.
Since in general, and in particular in large networks, Eq.~\ref{eq:noise-master-eq} is too difficult to solve directly, we focus on obtaining equations for first two momenta of $p(\vs)$.
Starting from the master equation one can derive the following self-consistency equations for the mean activity of units in a homogeneous network by assuming fluctuations around their mean input to be statistically independent \citep{helias2014intricorr}:
\begin{align*}
  \partial_t \langle s_i \rangle + \langle s_i \rangle = \frac{1}{2}\erfc{-\frac{\mu_i + b_i}{\sqrt{2}\sigma_i}}
\end{align*}
where the $\mu_i$ and $\sigma_i$ are given by Eq.~\ref{eq:supp-mu-network} and Eq.~\ref{eq:supp-sigma-network}, respectively.
To obtain the average activity in the stationary state, i.e., for $\partial_t \langle s_i \rangle = 0$, this equation needs to be solved self-consistently since the activity of unit $i$ can influence its input statistics ($\mu_i, \sigma_i$) through the recurrent connections.
By assuming homogeneous excitatory and inhibitory populations, the $\N$ dimensional problem reduces to a two-dimensional one \citep{helias2014intricorr}:
\begin{align}
  \langle s_\alpha \rangle = \frac{1}{2}\erfc{-\frac{\mu_\alpha + b_\alpha}{\sqrt{2}\sigma_\alpha}}
  \label{eq:supp-mf-activity}
\end{align}
with $\alpha \in \{\mathcal{E}, \mathcal{I}\}$.
The population-averaged equations for the mean and variance of the input hence are \citep{helias2014intricorr}:
\begin{align}
  \mu_\alpha =& \sum_\beta K_{\alpha \beta} w_{\alpha \beta} s_\beta \, ,
  \label{eq:supp-mf-popmean} \\
  \sigma^2_\alpha =& \sum_\beta K_{\alpha \beta} w_{\alpha \beta}^2 a_\beta + \sum_{\beta,\gamma}(K w)_{\alpha \beta}(K w)_{\alpha \gamma} c_{\beta\gamma} \, ,
  \label{eq:supp-mf-popvariance}
\end{align}
with $K_{EE}=K_{IE}=\gamma\N$, $K_{EI}=K_{II}=(1-\gamma)\N$ and $w_{EE}=w_{IE}=w$, $w_{EI}=w_{II}=-gw$.
To derive a self-consistency equation for pairwise correlations from the master equation one linearize the threshold activation function by considering a Gaussian distribution of the input field caused by recurrent inputs.
This leads to the following set of linear equations for the population-averaged covariances \citep{helias2014intricorr}:
\begin{align}
  2c_{\alpha\beta} = \sum_{\gamma}(\tilde{w}_{\alpha\gamma}c_{\gamma\beta}+\tilde{w}_{\beta\gamma}c_{\gamma\alpha})+\tilde{w}_{\alpha\beta}\frac{a_\beta}{N_\beta}+\tilde{w}_{\beta\alpha}\frac{a_\alpha}{N_\alpha} \;,
  \label{eq:supp-mf-corr}
\end{align}
with
\begin{align*}
  c_{\beta \gamma}=
  \begin{cases}
    \frac{1}{N_\beta (N_\beta - 1)}\sum_{i,j \in \beta, i\ne j}c_{ij} & \text{if } \beta = \gamma \\
    \frac{1}{N_\beta N_\gamma}\sum_{i \in \beta, j \in \gamma}c_{ij} & \text{else}
  \end{cases}
\end{align*}
The effective population-averaged weights $\tilde{w}_{\alpha \beta}$ are defined as:
\begin{align*}
  \tilde{w}_{\alpha\beta} := S(\mu_\alpha,\sigma_\alpha)K_{\alpha \beta}w_{\alpha \beta} \;,
\end{align*}
with the susceptibility given by $S(\mu_\alpha\sigma_\alpha) := \frac{1}{\sqrt{2\pi}\sigma_\alpha}\exp\left(-\frac{(\mu_\alpha+b_\alpha)^2}{2\sigma_\alpha}\right)$ \citep{helias2014intricorr}.
Since the average activity and covariances are mutually dependent, we employ an iterative numerical scheme in which we first determine the stationary activity under the assumption of zero correlations according to Eq.~\ref{eq:supp-mf-activity}.
Using this result we compute the population-average covariances from Eq.~\ref{eq:supp-mf-corr} which in turn can be used to improve the estimate for the stationary activity since they influence input statistics according to Eq.~\ref{eq:supp-mf-popvariance}.
We repeat this procedure until the values for population-averaged activities and covariances in two subsequent iterations do not differ significantly any more.
The mean activity and correlations in the recurrent background population obtained via this procedure, allows us to compute the input statistics in the sampling network and hence relate the inverse temperature to the mean and variance of the input as in Sec.~\ref{sec:supp-private-noise}.

Certain assumptions enter this analytical description of network statistics, which might not be fulfilled in general.
The description becomes much more complicated for spiking neuron models with non-linear subthreshhold dynamics like conductance-based neurons in neuromorphic systems \citep{petrovici2016stochastic}.
In this case, one can resort to empirically measuring the input statistics for a single isolated neuron given a certain arrangement of background sources (cf.~Sec.~\ref{sec:supp-calib-spiking}).
An advantage of this methods is that it is easy and straight forward to implement and will work for any configuration of background populations and sampling networks, allowing for arbitrary neuron models and parameters.
However, to estimate the statistics of the input accurately, one needs to collect statistics over a significant amount of time.

\subsection{Calibration (binary networks)}
The methods discussed above allow us to compute effective inverse temperature $\beta_\text{eff}$ from the statistics of different background inputs, either additive Gaussian noise, a population of intrinsically stochastic units or a recurrent network of deterministic units.
To approximate Boltzmann distributions via samples generated by networks with noise implemented via these alternative methods, we match their (effective) inverse temperatures.
A straightforward option is to adjust the noise parameter according to the desired input statistics.
While this is possible in the case of additive Gaussian noise for which we can freely adjust $\mu_i$ and $\sigma_i$, it is difficult to achieve in practice for the other methods.
We can achieve the same effect by rescaling the weights and biases in the sampling network.
The inverse temperature $\beta$ appears as a multiplicative factor in front of weights and biases in the stationary distribution of network states (Eq.~\ref{eq:supp-bm-dist}).
Scaling $\beta$ is hence equivalent to scaling all weights and biases by the inverse factor (see also \citep{grytskyy2013unified,dahmen16corrfluc,petrovici2016stochastic}).
An infinite amount of Boltzmann machines hence exists, all differing in weights ($w \rightarrow \alpha w, \alpha \in \mathbb{R}^+$), biases ($b \rightarrow \alpha b$) and inverse temperatures ($\beta \rightarrow \beta / \alpha$), producing statistically identical samples.
Given a mean background input $\mu_i$ and an effective inverse temperature $\beta_\text{eff}(\sigma_i)$ (cf.~Eq.~\ref{eq:supp-sigma-beta-integral}) arising from a particular realization of noise sources, we can emulate a Boltzmann machine at inverse temperature $\beta$ by rescaling all weights and biases in the sampling network according to
\begin{align}
  b_i \rightarrow& \beta / \beta_\text{eff} \, b_i - \mu_i \, , \\
  w_{ij} \rightarrow& \beta / \beta_\text{eff} \, w_{ij} \, .
\end{align}
This method hence only requires us to adapt weights and biases globally in the sampling network according to the statistics arising from an arbitrary realization of background input.

\subsection{Handwritten-digit generation}
\label{sec:supp-handwritten-digit}

In the generative task, we measure how well a sampling network with various realizations of background noise can approximate a trained data distribution, in contrast to the random distributions considered in the other simulations.
We use contrastive divergence (CD-$1$, \citep{hinton2002training}) to train a Boltzmann machine on a specific dataset.
We consider a dataset consisting of a subset of MNIST digits \citep{lecun1998mnist}, downscaled to $12$x$12$ pixels and with grayscale values converted to black and white.
We select one representative from each class ($0\dots9$) and extend the $144$ array determining the pixel values with $10$ entries for a one-hot encoding of the corresponding class, e.g., for the pattern zero, the last ten entries contain a $1$ at the first place and zeros otherwise.
These ten $154$ dimensional patterns form the prototype dataset.
A (noisy) training sample is generated by flipping every pixel from the first $144$ entries of a prototype pattern with probability $p^\text{flip}$.
After training, the network should represent a particular distribution $q^*$ over classes.
Training directly on a samples generated according to the class distribution $q^*$ will, in general, lead to a different stationary distribution of one-hot readout states $p$ generated by the network, since some patterns are more salient then others.
For example, by training on equal amounts of patterns of zeros and ones, the network will typically generate more zero states.
To nevertheless represent $q^*$ with the network, we iteratively train the Boltzmann machine choosing images and labels from a distribution $q$ that is adjusted between training sessions (Alg.~\ref{alg:noise-bm-training}).
Over many repetitions this procedure will lead to a stationary distribution of classes $p$ that closely approximates $q^*$.

\begin{algorithm}
  \begin{algorithmic}
    \State BM $\gets$ Boltzmann machine with random $w,b$
    \State $q \gets q^*$
    \For{$n_\text{it}$ iterations}
    \State collect $n_\text{sm}$ samples of label-unit states from BM
    \State $p \gets$ marginal distribution over one-hot label states
    \State $q \gets \lfloor (1-\alpha) q + \alpha q^* + \beta (q^* - p) \rfloor_+$ and normalize
    \State $d \gets$ generate training samples with class frequencies $q$
    \State train BM with CD-$1$ on dataset $d$ over $n_\text{ep}$ episodes
    \EndFor
  \end{algorithmic}
  \caption{Training of a fully visible Boltzmann machine via CD-$1$ to represent a particular distribution $q^*$ over label units with one-hot encoding.}
  \label{alg:noise-bm-training}
\end{algorithm}

After training a Boltzmann machine using this approach, we obtain a set of parameters, $w$ and $b$, that can be translated to parameters for sampling networks by appropriate rescaling as discussed above.
We collect samples by running the network in the absence of any input and recording the states of all label units.

\subsection{Calibration (spiking networks)}
\label{sec:supp-calib-spiking}

Similar as for binary units, we need to match the parameters for spiking sampling networks to their respective counterparts in Boltzmann machines.
Similar to \citep{petrovici2016stochastic} we use high-rate excitatory and inhibitory inputs to turn the deterministic behavior of a leaky-integrate-and-fire neuron into an effectively stochastic response.
However, in contrast to the original publication, we consider current based synapses for simplicity.
Since the calibration is performed on single cell level, we use the identical calibration scheme for the \private{}, \shared{} and \network{} case.
For a given configuration of noise sources, we first simulate the noise network with the specified parameters and measure its average firing rate.
The corresponding independent Poisson sources are set to fire with the same rate to ensure comparability between the two approaches.
The calibrations are then performed by varying the resting potential and recording the average activity of a single cell that is supplied with input from either a noise \network{} or Poisson sources. The \private{} case is calibrated separately in a similar manner.
By fitting the logistic function to the activation obtained by this procedure, we obtain two parameters, a shift and a scaling parameter, which are used to translate the synaptic weights from binary units to spiking neurons.
See \citep{petrovici2016stochastic} for details of the translation between the two domains.

%% file: supplement-supp.tex
\subsection{Pairwise input correlations}

Here we show how the covariance $C_{kl}^\text{in}$ between the input fields of two units $k$ and $l$ receiving inputs from a pool of sources can be decomposed into a part arising from shared inputs and another from activity correlations.
The input field for a single unit is given by Eq.~\ref{eq:supp-input-field} and hence:

\begin{align*}
  C_{kl}^\text{in} =& \EW{h_k h_l} - \EW{h_k}\EW{h_l} \\
  =& \EW{\left(\sum_i^K w_{ki} s_i + b_k\right) \left(\sum_j^K w_{lj}s_j + b_l\right)} - \EW{h_k}\EW{h_l} \\
  =& \sum_i^K w_{ki} w_{li} A_i + \sum_i^K \sum_{j\ne{}i}^K w_{ki} w_{lj} C_{ij} \\
  =& C^\text{in}_{\text{shared},kl} + C^\text{in}_{\text{corr},kl} \, .
\end{align*}
We introduced the auto- and crosscovariances $A_i=\EW{s_i^2}-\EW{s_i}^2$ and $C_{ij}=\EW{s_i s_j} - \EW{s_i}\EW{s_j}$ of the activities $s_i$ and $s_j$ of presynaptic neurons $i$ and $j$, respectively.
If Dale's law is respected and the sign of all outgoing connections of a sources is unique, i.e.~$\sgn{w_{ki}}=\sgn{w_{li}}$, $\forall k,l,i$, the first term is always positive ($C^\text{in}_{\text{shared},kl}>0$).
For a pool of independently active presynaptic neurons, $C_{ij}=0$ by definition and the second term in the input correlations hence vanishes ($C^\text{in}_{\text{corr},kl}=0$ $\forall k,l$).
The total input correlation is therefore always positive and determined by the number of shared sources.
If the presynaptic sources are units in a recurrently connected network, their pairwise correlation is in general non-zero ($C_{ij} \neq 0$).
In particular, in sparsely connected networks with sufficient inhibition, correlations arrange such that $C^\text{in}_{\text{corr},kl}\approx-C^\text{in}_{\text{shared},kl}$, leading to small remaining pairwise input correlations, $C_{kl}^\text{in}\approx 0$ \citep{renart2010asynchronous,tetzlaff2012decorrelation}.

\subsection{Sampling error depends on number of noise inputs per sampling unit}

To closely approximate the effect of Gaussian noise on the input field, one need a large number $\K$ of background inputs per sampling unit.
Here, we scale number of noise sources $\K$ per sampling unit, while also scaling the total number $\N$ of noise sources to keep their ratio constant.
This allows us to investigate the impact of $\K$ without altering the amount of shared-input correlations.
In addition to the three cases considered in main manuscript (private, shared, network noise), we additionally consider the case of a separate pool of noise sources for each sampling unit (``discrete''), where shared-input correlations are absent.

\begin{figure}[t]
  \centering
  \includegraphics[width=1.\linewidth]{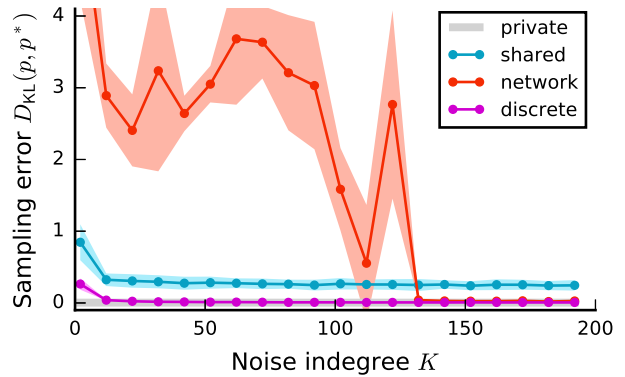}
  \caption{
    Sampling error $\DKL(p,p^*)$ as a function of the number of background inputs $\K$ per sampling unit.
    Error bands indicate mean $\pm$ SEM over $\valuetrials$ random network realizations.
    Magenta (``discrete'') uses $\K$ separate sources for each sampling unit.
    Sampling duration $T=\valueT\ms$.
    Connectivity constant $\K/\N = 0.9$.
    Remaining parameters as in Fig.~\ref{fig:results_convergence}.
  }
  \label{fig:supp-results_dkl_over_K}
\end{figure}

For small $\K$, the input distribution is strongly discretized and does not approximate Gaussian well, reflected in a large sampling error for very small $\K$ for the discrete and shared case (Fig.~\ref{fig:supp-results_dkl_over_K}).
As we increase $\K$, the sampling error decreases rapidly for the discrete case, and drops to the same level as Gaussian noise at about $50$ inputs.
For the shared case, the error decreases as well as we increase $\K$, but is limited from below by sampling error introduced by shared-input correlations.
For the network case, the sampling error is very large for small $\K$ as the network dynamics lock into a fixed point.
However, for $\K > 130$, the sampling error for the network case drops almost to the level of Gaussian noise.

\subsection{Small, recurrent networks can supply large sampling networks with noise -- no weight scaling}

In Fig.~\ref{fig:results_dkl_over_M} we scaled the weights in the sampling network with the size $\Nbm$ of the sampling network as $\weightscaling$.
Ignoring the influence of cross-correlations, this scaling keeps the variances of the input distribution arising from recurrent connections in the sampling network constant.
Effectively this leads to approximately constant entropy for a large range of sampling network sizes.

\begin{figure}[t]
  \centering
  \includegraphics[width=1.\linewidth]{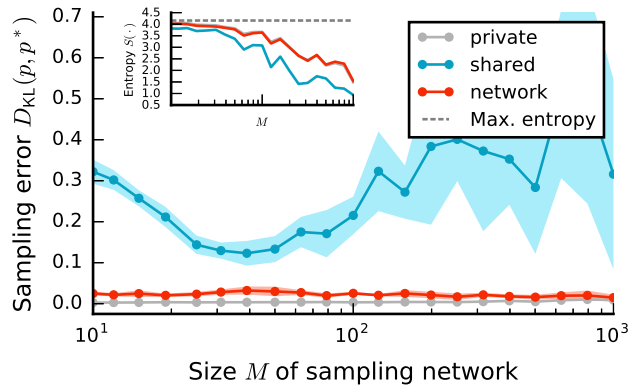} 
  \caption{
    Same as Fig.~\ref{fig:results_dkl_over_M}, but with constant average weight in sampling networks: $\mu_\text{BM}=-0.15$.
  }
  \label{fig:supp-results_dkl_over_M}
\end{figure}

If we do not scale the weights as above when increasing the size of the sampling network, the input variance increases and the relative noise strength hence decreases, leading to an effectively stronger coupled sampling network.
This strongly decreases the entropy of the sampled distribution (Fig.~\ref{fig:supp-results_dkl_over_M}, inset).
Despite the decrease in entropy, the sampling errors for the private and network cases stay approximately constant (Fig.~\ref{fig:supp-results_dkl_over_M}).
For the shared case, the sampling error initially decreases due to the strengthened effective feedback that suppresses shared-input correlations arising from the limited pool of background sources (cf.~{\bf Small, recurrent networks can supply large sampling networks with noise}).
As the size of the sampling network increases the sampling error increases again from about $\Nbm=40$.
This is most likely caused by the decrease in the relative noise strength and the sampling dynamics hence becoming too slow to approximate the target distribution in the finite sampling duration considered here.

\subsection{Simulation details}

Tab.~\ref{tab:supp-nordlie-binary-binary}, \ref{tab:supp-nordlie-binary-private}, \ref{tab:supp-nordlie-binary-shared}, \ref{tab:supp-nordlie-binary-network}, \ref{tab:supp-binary-params} summarize the binary network model and parameters.

Tab.~\ref{tab:supp-nordlie-spiking-private}, \ref{tab:supp-nordlie-spiking-shared}, \ref{tab:supp-nordlie-spiking-network}, \ref{tab:supp-spiking-params} summarize the spiking network model and parameters. Simulations carried out with NEST 2.10 \citep{bos2015nest}.

\begin{table*}

\setlength{\columnwidthleft}{0.18\textwidth}
\setlength{\columnwidthmiddle}{0.18\textwidth}

\begin{tabularx}{\textwidth}{|p{\columnwidthleft}|X|}
  \hline\modelhdr{2}{A}{Model summary}\\ \hline
  Populations    & One \\ \hline
  Topology       & \textemdash \\ \hline
  Connectivity   & All-to-all \\ \hline
  Neuron model   & Stochastic binary units \\ \hline
  Channel models & \textemdash \\ \hline
  Synapse model  & \textemdash \\ \hline
  Plasticity     & \textemdash \\ \hline
  External input & \textemdash \\ \hline
  Measurements   & Binary states of $\Nbmrec$ units \\ \hline
\end{tabularx} \\

\begin{tabularx}{\textwidth}{|p{\columnwidthleft}|p{\columnwidthmiddle}|X|}
  \hline\modelhdr{3}{B}{Populations}\\ \hline
  \bf Name & \bf Elements & \bf Size \\ \hline
  Sampling network & Stoch. binary units & $\Nbm$ \\ \hline
\end{tabularx} \\

\begin{tabularx}{\textwidth}{|p{\columnwidthleft}|p{\columnwidthmiddle}|X|}
  \hline\modelhdr{3}{C}{Connectivity}\\ \hline
  \bf Source & \bf Target & \bf Pattern \\ \hline
  Sampling network & Sampling network & All-to-all, random weights drawn from Beta distribution, $w_{ij} \sim \text{Beta}(a, b)$,
  symmetric connections $w_{ij}=w_{ji}$, no self connections $w_{ii} = 0 $ \\ \hline
\end{tabularx} \\

\begin{tabularx}{\textwidth}{|p{\columnwidthleft}|X|}
  \hline\modelhdr{2}{D}{Neuron model}\\ \hline
  Type & Stochastic binary units \\
  Dynamics & Transition into state $1$ according to probability determined by the activation function $F_i(h_i ) = \Fstoch{h_i}$ with input field $h_i = \sum_j w_{ij}s_j + b_i$. \\ \hline
\end{tabularx}

\caption{
  Description of the sampling network model with \intrinsic{} noise (according to \citep{nordlie2009towards}).
  \label{tab:supp-nordlie-binary-binary}
}

\end{table*}

\begin{table*}

\setlength{\columnwidthleft}{0.18\textwidth}
\setlength{\columnwidthmiddle}{0.18\textwidth}

\begin{tabularx}{\textwidth}{|p{\columnwidthleft}|X|}
  \hline\modelhdr{2}{A}{Model summary}\\ \hline
  Populations    & One \\ \hline
  Topology       & \textemdash \\ \hline
  Connectivity   & All-to-all \\ \hline
  Neuron model   & Stochastic binary units \\ \hline
  Channel models & \textemdash \\ \hline
  Synapse model  & \textemdash \\ \hline
  Plasticity     & \textemdash \\ \hline
  External input & \textemdash \\ \hline
  Measurements   & Binary states of $\Nbmrec$ units \\ \hline
\end{tabularx} \\

\begin{tabularx}{\textwidth}{|p{\columnwidthleft}|p{\columnwidthmiddle}|X|}
  \hline\modelhdr{3}{B}{Populations}\\ \hline
  \bf Name & \bf Elements & \bf Size \\ \hline
  Sampling network & Stoch. binary units & $\Nbm$ \\ \hline
\end{tabularx} \\

\begin{tabularx}{\textwidth}{|p{\columnwidthleft}|p{\columnwidthmiddle}|X|}
  \hline\modelhdr{3}{C}{Connectivity}\\ \hline
  \bf Source & \bf Target & \bf Pattern \\ \hline
  Sampling network & Sampling network & All-to-all, random weights drawn from Beta distribution, $w_{ij} \sim \text{Beta}(a, b)$,
  symmetric connections $w_{ij}=w_{ji}$, no self connections $w_{ii} = 0 $ \\ \hline
\end{tabularx} \\

\begin{tabularx}{\textwidth}{|p{\columnwidthleft}|X|}
  \hline\modelhdr{2}{D}{Neuron model}\\ \hline
  Type & Stochastic binary units \\
  Dynamics & Transition into state $1$ according to probability determined by the activation function $F_i(h_i) = \Ferfc{h_i + \mu_i}$ with input field $h_i = \sum_j w_{ij}s_j + b_i$. \\ \hline
\end{tabularx}

\caption{
  Description of sampling network model with \private{} noise (according to \citep{nordlie2009towards}).
  \label{tab:supp-nordlie-binary-private}
}

\end{table*}

\begin{table*}

\setlength{\columnwidthleft}{0.18\textwidth}
\setlength{\columnwidthmiddle}{0.18\textwidth}

\begin{tabularx}{\textwidth}{|p{\columnwidthleft}|X|}
  \hline\modelhdr{2}{A}{Model summary}\\ \hline
  Populations    & Three \\ \hline
  Topology       & \textemdash \\ \hline
  Connectivity   & All-to-all; sparse random with fixed indegree \\ \hline
  Neuron model   & Stochastic binary units, deterministic binary units \\ \hline
  Channel models & \textemdash \\ \hline
  Synapse model  & \textemdash \\ \hline
  Plasticity     & \textemdash \\ \hline
  External input & \textemdash \\ \hline
  Measurements   & Binary states \\ \hline
\end{tabularx} \\

\begin{tabularx}{\textwidth}{|p{\columnwidthleft}|p{\columnwidthmiddle}|X|}
  \hline\modelhdr{3}{B}{Populations}\\ \hline
  \bf Name & \bf Elements & \bf Size \\ \hline
  Sampling network & Det. binary units & $\Nbm$ \\ \hline
  Background pop. (E) & Stoch. binary units & $\gamma \N$ \\ \hline
  Background pop. (I) & Stoch. binary units & $(1 - \gamma) \N$ \\ \hline
\end{tabularx} \\

\begin{tabularx}{\textwidth}{|p{\columnwidthleft}|p{\columnwidthmiddle}|X|}
  \hline\modelhdr{3}{C}{Connectivity}\\ \hline
  \bf Source & \bf Target & \bf Pattern \\ \hline
  Sampling network & Sampling network & All-to-all, random weights drawn from Beta distribution, $w_{ij} \sim \text{Beta}(a, b)$,
  symmetric connections $w_{ij}=w_{ji}$, no self connections $w_{ii} = 0 $ \\ \hline
  Background pop. (E) & Sampling network & Random convergent $\gamma\K \rightarrow 1$, weight $w$ \\ \hline
  Background pop. (I) & Sampling network & Random convergent $(1-\gamma)\K \rightarrow 1$, weight $-gw$ \\ \hline
\end{tabularx} \\

\begin{tabularx}{\textwidth}{|p{\columnwidthleft}|X|}
  \hline\modelhdr{2}{D}{Neuron model}\\ \hline
  Type & Stochastic binary units \\
  Dynamics & Transition into state $1$ according to probability determined by the activation function $F_i(h_i ) = \Fstoch{h_i}$ with input field $h_i = \sum_j w_{ij}s_j + b_i$. \\ \hline
  Type & Deterministic binary units \\
  Dynamics & Transition into state $1$ according to the activation function $F_i(h_i) = \Theta(h_i)$ with input field $h_i = \sum_j w_{ij}s_j + b_i$. \\ \hline
\end{tabularx}

\begin{tabularx}{\textwidth}{|X|}
  \hline\modelhdr{1}{E}{Measurements}\\ \hline
  Binary states of $\Nbmrec$ units from sampling network \\ \hline
\end{tabularx}

\caption{
  Description of sampling network model with \shared{} noise (according to \citep{nordlie2009towards}).
  \label{tab:supp-nordlie-binary-shared}
}

\end{table*}

\begin{table*}

\setlength{\columnwidthleft}{0.18\textwidth}
\setlength{\columnwidthmiddle}{0.18\textwidth}

\begin{tabularx}{\textwidth}{|p{\columnwidthleft}|X|}
  \hline\modelhdr{2}{A}{Model summary}\\ \hline
  Populations    & Three \\ \hline
  Topology       & \textemdash \\ \hline
  Connectivity   & All-to-all; sparse random with fixed indegree \\ \hline
  Neuron model   & Deterministic binary units \\ \hline
  Channel models & \textemdash \\ \hline
  Synapse model  & \textemdash \\ \hline
  Plasticity     & \textemdash \\ \hline
  External input & \textemdash \\ \hline
  Measurements   & Binary states \\ \hline
\end{tabularx} \\

\begin{tabularx}{\textwidth}{|p{\columnwidthleft}|p{\columnwidthmiddle}|X|}
  \hline\modelhdr{3}{B}{Populations}\\ \hline
  \bf Name & \bf Elements & \bf Size \\ \hline
  Sampling network & Det. binary units & $\Nbm$ \\ \hline
  Background pop. (E) & Det. binary units & $\gamma \N$ \\ \hline
  Background pop. (I) & Det. binary units & $(1 - \gamma) \N$ \\ \hline
\end{tabularx} \\

\begin{tabularx}{\textwidth}{|p{\columnwidthleft}|p{\columnwidthmiddle}|X|}
  \hline\modelhdr{3}{C}{Connectivity}\\ \hline
  \bf Source & \bf Target & \bf Pattern \\ \hline
  Sampling network & Sampling network & All-to-all, random weights drawn from Beta distribution, $w_{ij} \sim \text{Beta}(a, b)$,
  symmetric connections $w_{ij}=w_{ji}$, no self connections $w_{ii} = 0 $ \\ \hline
  Background pop. (E) & Sampling network & Random convergent $\gamma\K \rightarrow 1$, weight $w$ \\ \hline
  Background pop. (I) & Sampling network & Random convergent $(1-\gamma)\K \rightarrow 1$, weight $-gw$ \\ \hline
  Background pop. (E) & Background pop. (E) & Random convergent $\gamma\K \rightarrow 1$, weight $w$ \\ \hline
  Background pop. (E) & Background pop. (I) & Random convergent $\gamma\K \rightarrow 1$, weight $w$ \\ \hline
  Background pop. (I) & Background pop. (E) & Random convergent $(1-\gamma)\K \rightarrow 1$, weight $-gw$ \\ \hline
  Background pop. (I) & Background pop. (I) & Random convergent $(1-\gamma)\K \rightarrow 1$, weight $-gw$ \\ \hline
\end{tabularx} \\

\begin{tabularx}{\textwidth}{|p{\columnwidthleft}|X|}
  \hline\modelhdr{2}{D}{Neuron model}\\ \hline
  Type & Deterministic binary units \\
  Dynamics & Transition into state $1$ according to the activation function $F_i(h_i) = \Theta(h_i)$ with input field $h_i = \sum_j w_{ij}s_j + b_i$. \\ \hline
\end{tabularx}

\begin{tabularx}{\textwidth}{|X|}
  \hline\modelhdr{1}{E}{Measurements}\\ \hline
  Binary states of $\Nbmrec$ units from sampling network \\ \hline
\end{tabularx}

\caption{
  Description of sampling network model with \network{} noise (according to \citep{nordlie2009towards}).
  \label{tab:supp-nordlie-binary-network}
}

\end{table*}

\begin{table*}

\setlength{\columnwidthleft}{0.2\textwidth}
\setlength{\columnwidthmiddle}{0.2\textwidth}

\begin{tabularx}{\textwidth}{|p{\columnwidthleft}|X|}
  \hline\parameterhdr{2}{B}{Populations}\\\hline
  \bf Name & \bf Values \\ \hline
  $\Nbm$ & $\valueNbm^*$ \\
  $\N$ & $\valueN^*$ \\
  $\gamma$ & $\valuegamma$ \\ \hline
\end{tabularx} \\

\begin{tabularx}{\textwidth}{|p{\columnwidthleft}|X|}
  \hline\parameterhdr{2}{C}{Connectivity}\\\hline
  \bf Name & \bf Values \\ \hline
  $a$ & $\valuea$ \\
  $b$ & $\valueb$ \\
  $\K$ & $\valueK$ \\
  $w$ & $\valuew$ \\
  $g$ & $\valueg$ \\ \hline
\end{tabularx} \\

\begin{tabularx}{\textwidth}{|p{\columnwidthleft}|X|}
  \hline\parameterhdr{2}{D}{Neuron model}\\\hline
  \bf Name & \bf Values \\ \hline
  $\beta$ & $\valuebeta^*$ \\
  $\mu$ & $\valuemu$ \\
  $\sigma$ & from $\beta$ via Eq.~\ref{eq:supp-sigma-beta-integral} \\ \hline
\end{tabularx} \\

\begin{tabularx}{\textwidth}{|p{\columnwidthleft}|X|}
  \hline\parameterhdr{2}{E}{Measurements}\\\hline
  \bf Name & \bf Values \\ \hline
  $\Nbmrec$ & $\valueNbmrec$ \\ \hline
\end{tabularx} \\

\begin{tabularx}{\textwidth}{|p{\columnwidthleft}|p{\columnwidthmiddle}|X|}
  \hline\parameterhdr{3}{}{Miscellaneous}\\\hline
  \bf Name & \bf Values & \bf Description \\ \hline
  $\bar{s}$ & $0.4$ & Average activity in sampling networks \\
  $\bar{z}$ & $0.3$ & Average activity in background population \\
  $T_\text{sim}$ & $\valueT\ms^*$ & Simulation time \\
  $T_\text{warmup}$ & $500\ms$ & Warmup time (ignored during analysis) \\ \hline
  $\tau$ & $10\ms$ & Average inter-update interval \\ \hline
\end{tabularx} \\

\caption{
    Parameters for binary network simulations (according to \citep{nordlie2009towards}). Stars indicate default values.
  \label{tab:supp-binary-params}
}

\end{table*}

\begin{table*}

\setlength{\columnwidthleft}{0.18\textwidth}
\setlength{\columnwidthmiddle}{0.18\textwidth}

\begin{tabularx}{\textwidth}{|p{\columnwidthleft}|X|}
  \hline\modelhdr{2}{A}{Model summary}\\\hline
  Populations    & One \\ \hline
  Topology       & \textemdash \\ \hline
  Connectivity   & All-to-all \\ \hline
  Neuron model   & Leaky integrate-and-fire (LIF) \\ \hline
  Channel models & \textemdash \\ \hline
  Synapse model  & Exponentially decaying currents, fixed delays \\ \hline
  Plasticity     & \textemdash \\ \hline
  External input & Poisson-distributed spike trains \\ \hline
  Measurements   & Spikes \\ \hline
\end{tabularx} \\

\begin{tabularx}{\textwidth}{|p{\columnwidthleft}|p{\columnwidthmiddle}|X|}
  \hline\modelhdr{3}{B}{Populations}\\\hline
  \bf Name & \bf Elements & \bf Size \\
  \hline
  Sampling network & LIF neuron & $\Nbm$ \\
  \hline
\end{tabularx} \\

\begin{tabularx}{\textwidth}{|p{\columnwidthleft}|p{\columnwidthmiddle}|X|}
  \hline\modelhdr{3}{C}{Connectivity}\\\hline
  \bf Source & \bf Target & \bf Pattern \\ \hline
  Sampling network & Sampling network & All-to-all, random weights drawn from Beta distribution, $w_{ij} \sim \text{Beta}(a, b)$, symmetric connections $w_{ij}=w_{ji}$, no self connections $w_{ii} = 0 $, %
  translation from binary-unit domain to spiking neurons via constant calibration factors (see Sec.~\ref{sec:supp-calib-spiking}) \\ \hline
  \hline
\end{tabularx} \\

\begin{tabularx}{\textwidth}{|p{\columnwidthleft}|X|}
  \hline\modelhdr{2}{D}{Neuron and synapse model}\\\hline
  Type & Leaky integrate-and-fire, exponential currents \\ \hline
  Subthreshold dynamics &
    Subthreshold dynamics ($t \not\in (t^*, t^* + \tauref)$): \newline
    \hspace*{1em} $\cmem \frac{\text{d}}{\text{d}t} \vm(t) = -\gleak (\vm(t) - \vrest) + \Isyn(t)$ \newline
    Reset and refractoriness ($t \in (t^*, t^* + \tauref)$): \newline
    \hspace*{1em} $\vm(t) = \vreset$ \\ \hline
  Current dynamics &
    \hspace*{1em} $\tausyn \frac{\text{d}}{\text{d}t} \Isyn(t) = -\Isyn(t) + \sum_{i,k} J \delta(t-t_i^k-d)$ \newline
    Here the sum over $i$ runs over all presynaptic neurons and the sum over $k$ over all spike times of the respective neuron $i$ \\ \hline
  Spiking &
    If $\vm(t^*-) < \vthresh \wedge \vm(t^*+) \ge \vthresh$: \newline
    \hspace*{1em} emit spike with time stamp $t^*$ \\ \hline
\end{tabularx}

\begin{tabularx}{\textwidth}{|X|}
  \hline\modelhdr{1}{E}{Measurements}\\ \hline
  Spike trains recorded from $\Nbmrec$ neurons from the sampling network \\ \hline
\end{tabularx}

\begin{tabularx}{\textwidth}{|X|}
  \hline\modelhdr{1}{F}{External input}\\ \hline
  Per neuron, one private excitatory and one inhibitory Poisson source with rate $\nuex$ and $\nuin$, respectively. \\ \hline
\end{tabularx}

\caption{
  Description of spiking sampling network model with \private{} noise (according to \citep{nordlie2009towards}).
  \label{tab:supp-nordlie-spiking-private}
}

\end{table*}

\begin{table*}

\setlength{\columnwidthleft}{0.18\textwidth}
\setlength{\columnwidthmiddle}{0.18\textwidth}

\begin{tabularx}{\textwidth}{|p{\columnwidthleft}|X|}
  \hline\modelhdr{2}{A}{Model summary}\\\hline
  Populations    & One \\ \hline
  Topology       & \textemdash \\ \hline
  Connectivity   & All-to-all; sparse random with fixed indegree \\ \hline
  Neuron model   & Leaky integrate-and-fire (LIF) \\ \hline
  Channel models & \textemdash \\ \hline
  Synapse model  & Exponentially decaying currents, fixed delays \\ \hline
  Plasticity     & \textemdash \\ \hline
  External input & Poisson-distributed spike trains \\ \hline
  Measurements   & Spikes \\ \hline
\end{tabularx} \\

\begin{tabularx}{\textwidth}{|p{\columnwidthleft}|p{\columnwidthmiddle}|X|}
  \hline\modelhdr{3}{B}{Populations}\\\hline
  \bf Name & \bf Elements & \bf Size \\
  \hline
  Sampling network & LIF neuron & $\Nbm$ \\
  \hline
\end{tabularx} \\

\begin{tabularx}{\textwidth}{|p{\columnwidthleft}|p{\columnwidthmiddle}|X|}
  \hline\modelhdr{3}{C}{Connectivity}\\\hline
  \bf Source & \bf Target & \bf Pattern \\ \hline
  Sampling network & Sampling network & All-to-all, random weights drawn from Beta distribution, $w_{ij} \sim \text{Beta}(a, b)$, symmetric connections $w_{ij}=w_{ji}$, no self connections $w_{ii} = 0 $, %
  translation from binary-unit domain to spiking neurons via constant calibration factors (see Sec.~\ref{sec:supp-calib-spiking}) \\
  \hline
\end{tabularx} \\

\begin{tabularx}{\textwidth}{|X|}
  \hline\modelhdr{1}{D}{Neuron and synapse model}\\\hline
  %
  See Tab.~\ref{tab:supp-nordlie-spiking-private}. \\ \hline
\end{tabularx}

\begin{tabularx}{\textwidth}{|X|}
  \hline\modelhdr{1}{E}{Measurements}\\ \hline
  See Tab.~\ref{tab:supp-nordlie-spiking-private}. \\ \hline
\end{tabularx}

\begin{tabularx}{\textwidth}{|X|}
  \hline\modelhdr{1}{F}{External input}\\ \hline
  Per neuron, $\gamma\K$ excitatory and $(1-\gamma)\K$ inhibitory Poisson sources with weight $J$, rate $\tilde{\nu}_\text{ex}$ and weight $-gJ$, rate $\tilde{\nu}_\text{in}$, respectively. Excitatory and inhibitory inputs randomly chosen from a common pool of $\gamma \N$ and $(1-\gamma)\N$ units, respectively. \\ \hline
\end{tabularx}

\caption{%
  Description of spiking sampling network model with \shared{} noise (according to \citep{nordlie2009towards}).
  \label{tab:supp-nordlie-spiking-shared}
}

\end{table*}

\begin{table*}

\setlength{\columnwidthleft}{0.18\textwidth}
\setlength{\columnwidthmiddle}{0.18\textwidth}

\begin{tabularx}{\textwidth}{|p{\columnwidthleft}|X|}
  \hline\modelhdr{2}{A}{Model summary}\\\hline
  Populations    & Three \\ \hline
  Topology       & \textemdash \\ \hline
  Connectivity   & All-to-all; sparse random with fixed indegree \\ \hline
  Neuron model   & Leaky integrate-and-fire (LIF) \\ \hline
  Channel models & \textemdash \\ \hline
  Synapse model  & Exponentially decaying currents, fixed delays \\ \hline
  Plasticity     & \textemdash \\ \hline
  External input & Resting potential above firing threshold in background populations \\ \hline
  Measurements   & Spikes \\ \hline
\end{tabularx} \\

\begin{tabularx}{\textwidth}{|p{\columnwidthleft}|p{\columnwidthmiddle}|X|}
  \hline\modelhdr{3}{B}{Populations}\\\hline
  \bf Name & \bf Elements & \bf Size \\
  \hline
  Sampling network & LIF neuron & $\Nbm$ \\
  Background pop. (E) & LIF neuron & $\gamma \N$ \\
  Background pop. (I) & LIF neuron & $(1 - \gamma) \N$ \\
  \hline
\end{tabularx} \\

\begin{tabularx}{\textwidth}{|p{\columnwidthleft}|p{\columnwidthmiddle}|X|}
  \hline\modelhdr{3}{C}{Connectivity}\\\hline
  \bf Source & \bf Target & \bf Pattern \\ \hline
  Sampling network & Sampling network & All-to-all, random weights drawn from Beta distribution, $w_{ij} \sim \text{Beta}(a, b)$, symmetric connections $w_{ij}=w_{ji}$, no self connections $w_{ii} = 0 $, %
  translation from binary-unit domain to spiking neurons via constant calibration factors (see Sec.~\ref{sec:supp-calib-spiking}) \\
  Background pop. (E) & Sampling network & Random convergent, $\gamma \K \rightarrow 1$, weight $w$, delay $d$ \\
  Background pop. (I) & Sampling network & Random convergent, $(1 - \gamma) \K \rightarrow 1$, weight $-gw$, delay $d$ \\
  Background pop. (E) & Background pop. (E) & Random convergent, $\gamma \K \rightarrow 1$, weight $w$, delay $d$ \\
  Background pop. (E) & Background pop. (I) & Random convergent, $\gamma \K \rightarrow 1$, weight $w$, delay $d$ \\
  Background pop. (I) & Background pop. (E) & Random convergent, $(1 - \gamma) \K \rightarrow 1$, weight $-gw$, delay $d$ \\
  Background pop. (I) & Background pop. (I) & Random convergent, $(1 - \gamma) \K \rightarrow 1$, weight $-gw$, delay $d$ \\
  \hline
\end{tabularx} \\

\begin{tabularx}{\textwidth}{|X|}
  \hline\modelhdr{1}{D}{Neuron and synapse model}\\\hline
  See Tab.~\ref{tab:supp-nordlie-spiking-private}. \\ \hline
\end{tabularx}

\begin{tabularx}{\textwidth}{|X|}
  \hline\modelhdr{1}{E}{Measurements}\\ \hline
  See Tab.~\ref{tab:supp-nordlie-spiking-private}. \\ \hline
\end{tabularx}

\caption{
  Description of spiking sampling network model with \network{} noise (according to \citep{nordlie2009towards}).
  \label{tab:supp-nordlie-spiking-network}
}

\end{table*}

\begin{table*}

\setlength{\columnwidthleft}{0.2\textwidth}
\setlength{\columnwidthmiddle}{0.2\textwidth}

\begin{tabularx}{\textwidth}{|X|}
  \hline\parameterhdr{1}{B}{Populations}\\\hline
  See Tab.~\ref{tab:supp-binary-params}. \\ \hline
\end{tabularx} \\

\begin{tabularx}{\textwidth}{|p{\columnwidthleft}|X|}
  \hline\parameterhdr{2}{C}{Connectivity}\\\hline
  \bf Name & \bf Values \\ \hline
  $a$ & $\valuea$ \\
  $b$ & $\valueb$ \\
  $\K$ & $\valueKlif$ \\
  $J$ & $0.002\nA$ ($0.0635\nA$) \\
  $g$ & $2$ \\
  $d$ & $0.1\ms$ ($1.0\ms$) \\ \hline
\end{tabularx} \\

\begin{tabularx}{\textwidth}{|p{\columnwidthleft}|X|}
  \hline\parameterhdr{2}{D}{Neuron model}\\\hline
  \bf Name & \bf Values \\ \hline
  $\tauref$ & $10.0\ms$ ($0.1\ms$) \\
  $\tausyn$ & $10.0\ms$ ($5.0\ms$) \\
  $\cmem$ & $0.2\nF$ ($1.0\nF$) \\
  $\gleak$ & $2.0\uS$ ($0.05\uS$) \\
  $\vrest$ & $-50.00\mV$ ($-40.00\mV$) \\
  $\vreset$ & $-50.01\mV$ ($-60.00\mV$) \\
  $\vthresh$ & $-50.00\mV$ \\ \hline
\end{tabularx} \\

\begin{tabularx}{\textwidth}{|p{\columnwidthleft}|X|}
  \hline\parameterhdr{2}{E}{Measurements}\\\hline
  \bf Name & \bf Values \\ \hline
  $\Nbmrec$ & $\valueNbmreclif$ \\ \hline
\end{tabularx} \\

\begin{tabularx}{\textwidth}{|p{\columnwidthleft}|p{\columnwidthmiddle}|X|}
  \hline\parameterhdr{3}{}{Miscellaneous}\\\hline
  \bf Name & \bf Values & \bf Description \\ \hline
  $T_\text{sim}$ & $10^7\ms$ & Simulation time \\
  $T_\text{warmup}$ & $10^3\ms$ & Warmup time (ignored during analysis) \\ \hline
\end{tabularx} \\

\begin{tabularx}{\textwidth}{|p{\columnwidthleft}|X|}
  \hline\parameterhdr{2}{F}{External input}\\\hline
  \bf Name & \bf Values \\ \hline
  $\nuex$ & $10\kHz$ \\
  $\nuin$ & $10\kHz$ \\
  $\tilde{\nu}_\text{ex}$ & $4.4 \pm 0.1\Hz$ \\
  $\tilde{\nu}_\text{in}$ & $4.4 \pm 0.1\Hz$ \\ \hline
\end{tabularx} \\

\caption{
    Table of parameters for spiking network simulations (according to \citep{nordlie2009towards}). Values without parantheses are for the sampling network, values in parantheses for the noise network.
  \label{tab:supp-spiking-params}
}

\end{table*}